\documentclass[epsf,prb,reprint,twocolumn,showpacs,nofootinbib,superscriptaddress]{revtex4-1}

\usepackage[pdftex]{graphicx}
\usepackage{dcolumn}
\usepackage{bm} 
\usepackage{epsfig}
\usepackage{latexsym}
\usepackage{amsmath}
\usepackage{amsfonts}
\usepackage{amssymb}
\usepackage{color}
\usepackage{array}
\usepackage{framed}

\usepackage{hyperref}
\usepackage{graphicx}
\usepackage{enumitem}
\usepackage[T1]{fontenc}
\usepackage[normalem]{ulem}
\usepackage{lineno}
\usepackage{physics}
\usepackage[rawfloats=true]{floatrow} 
\usepackage[titletoc]{appendix}

\newcommand{\rtext}[1]{{\color{black}#1}}

\usepackage[normalem]{ulem}
\newcommand{\oldtext}[1]{{{\color{blue}\sout{}}}}

\begin{document}

\title{Fractons from polarons}

\author{John Sous}\email{Current address: Department of Physics, Columbia University, New York, New York 10027, USA}
\affiliation{Department of Physics T42, Technische Universit\"at M\"unchen, Garching, 85747, Germany} 
\affiliation{Department of Physics \& Astronomy, University of British Columbia, Vancouver, BC V6T 1Z3, Canada}

\author{Michael Pretko}
\affiliation{Department of Physics and Center for Theory of Quantum Matter, University of Colorado, Boulder, CO 80309, USA}

\date{\today}

\begin{abstract} 
Fractons are a type of emergent quasiparticle that cannot move freely in isolation, but can easily move in bound pairs. Similar phenomenology is found in boson-affected hopping models, encountered in the study of polaron systems and hole-doped Ising antiferromagnets, in which motion of a particle requires the creation or absorption of background bosonic excitations. In such models, individual low-energy quasiparticles cannot move freely, while bound pairs have drastically increased mobility. We show that boson-affected hopping models can provide a natural realization of fractons, either approximately or exactly, depending on the details of the system. We first consider a generic one-dimensional boson-affected hopping model, in which we show that single particles move only at sixth order in perturbation theory, while motion of bound states occurs at second order, allowing for a broad parameter regime exhibiting approximate fracton phenomenology. We explicitly map the model onto a fracton Hamiltonian featuring conservation of dipole moment via integrating out the mediating bosons. We then consider a special type of boson-affected hopping models with mutual hard-core repulsion between particles and bosons, experimentally accessible in hole-doped mixed-dimensional Ising antiferromagnets, in which the hole motion is one dimensional in an otherwise two-dimensional antiferromagnetic background. We show that this system exhibits {\em perfect} fracton behavior to all orders in perturbation theory. We suggest diagnostic signatures of fractonic behavior, opening a door to use already existing experimental tools to study their unusual physics, such as universal gravitation and restricted thermalization. As an example, gravitational attraction manifests as phase separation of holes in doped antiferromagnets. In studying these models, we identify simple effective one-dimensional microscopic Hamiltonians featuring perfect fractonic behavior, paving the way to future studies on fracton physics in lower dimensions, where a wealth of numerical and analytical tools already exist. In these Hamiltonians, we identify pair-hopping interactions as the mechanism of dipole motion, and argue that this may provide a connection to topological edge states in boundary fractonic systems.

\end{abstract}

\maketitle

\normalsize

\section{Introduction}

Strongly interacting quantum many-body systems can exist in phases of matter featuring a wide range of exotic phenomena, including topologically protected edge currents and quantum error-correcting properties.\cite{Nayak, Devitt}  Perhaps most surprisingly, such phases can exhibit exotic new quasiparticles resulting from fractionalization of the fundamental particles comprising the system.  For example, celebrated fractional quantum Hall systems host quasiparticles carrying only a fraction of the charge of an individual electron.\cite{Hansson} These fractionalized quasiparticles typically also feature anyonic statistics, partway between ordinary bosons and fermions.  Further types of fractionalization are also possible such as the separation of the spin and charge of the constituent electrons.\cite{spincharge1,spincharge2,MonaSpinCharge,EugeneSpinCharge}

\begin{figure}[!htb]
 \centering
  \includegraphics[width=0.45\columnwidth]{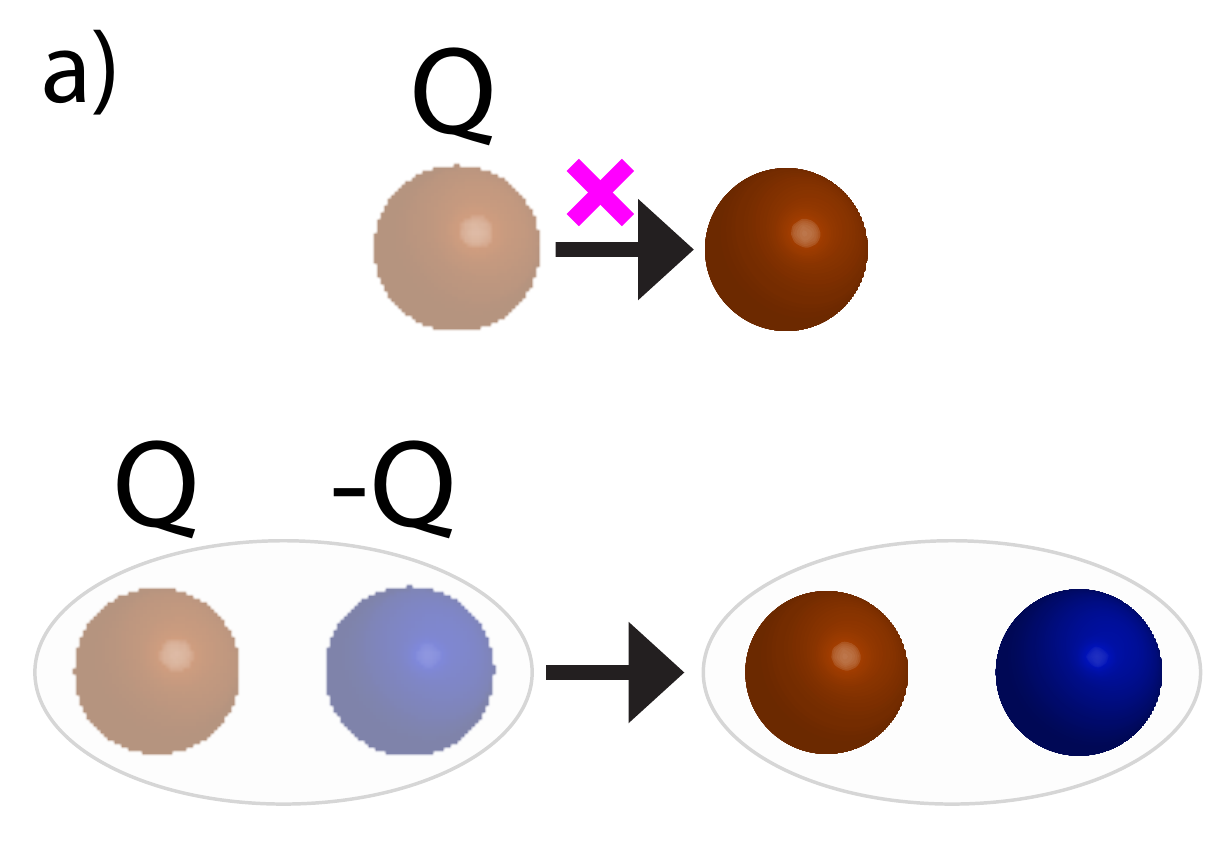} \includegraphics[width=0.45\columnwidth]{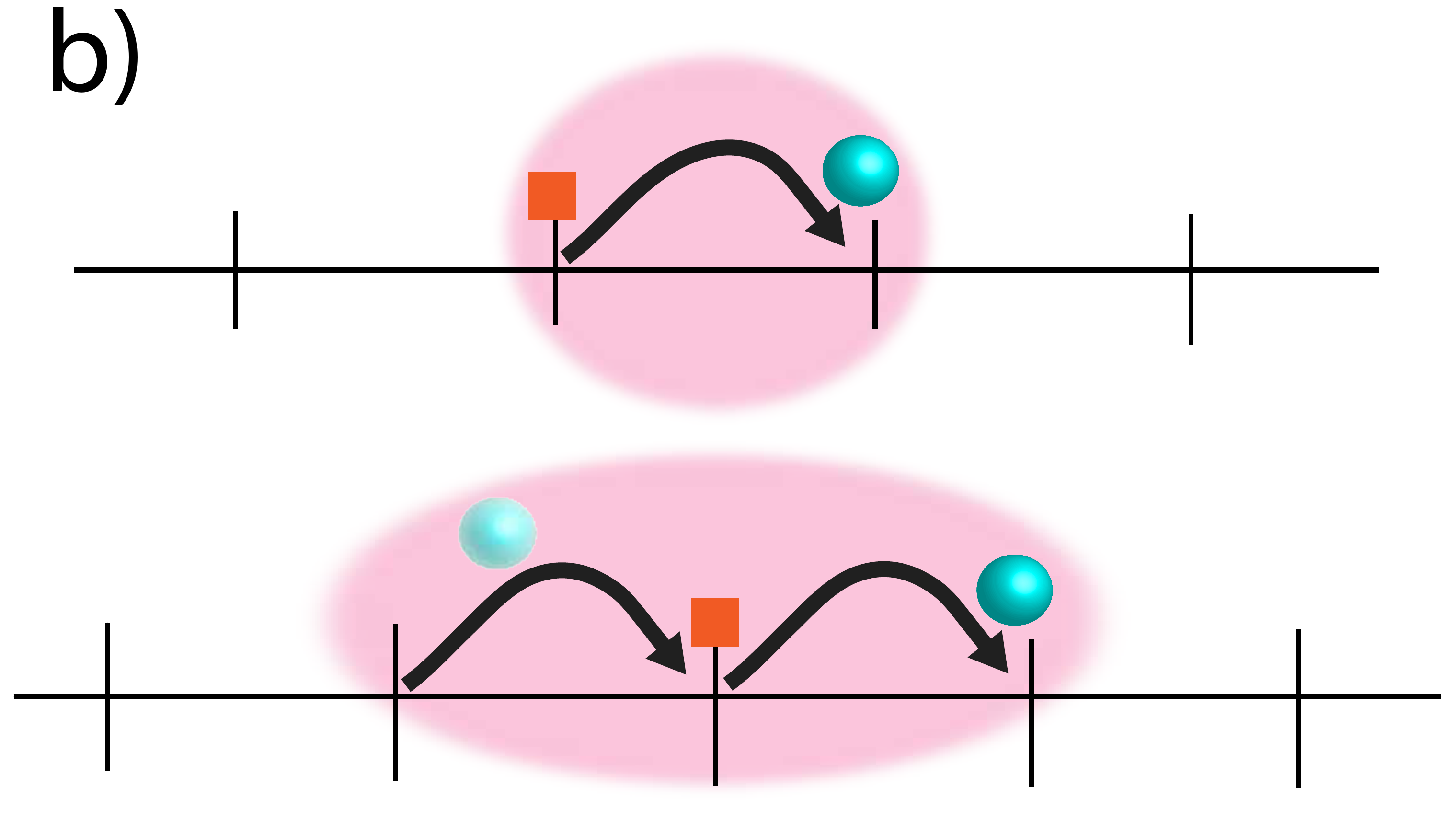} 
 \caption{{\bf{Fractons from polarons.}}
  Fractons and dipoles are schematically represented in a): A fracton of charge $\rm{Q}$ cannot move, but a bound state of two fractons (typically of opposite charge) can, in accordance with dipole conservation. Boson-affected hopping models, used to study certain polaron systems, are represented in b) where a particle (light-blue dot) can only move by creating a boson (orange square) at its departure site or by absorbing one at its arrival site.  The particle-bath coupling leads to formation of polarons: a particle dressed by a bosonic cloud (cloudy light-red oval).  A single particle (upper panel), in forming a polaron, becomes localized by string excitations, while two particles (lower panel) become bound (forming a bipolaron) and can move via boson-mediated pair-hopping interactions.  This behavior is exact in one-dimensional systems with mutual hard-core repulsion between the particles and bosons, and is approximate in absence of the constraint. By staggering charge, see Subsection~\ref{Staggering}, we identify a single polaron as a fracton, and a bipolaron as a dipole.
 }
 \label{fig:polaron}
 \end{figure}

While the notion of fractionalized quasiparticles has been understood for several decades, theoretical work over the past few years has uncovered a new more unusual type of quasiparticle.  Certain quantum phases of matter are known to host ``fracton'' quasiparticles, characterized by an exotic set of mobility restrictions.  Specifically, a fracton is a quasiparticle that does not have the ability to move by itself.  Rather, fractons can only move by coming together to form certain mobile bound states, see Figure~\ref{fig:polaron}. This restriction on mobility is usually encoded in the system in the form of higher-moment charge conservation laws, such as the conservation of dipole moment that often arises as a consequence of an emergent symmetric tensor gauge field.\cite{sub,genem,witten,higgs1,higgs2}

These unusual fracton quasiparticles have attracted intense interest due to their fundamental importance as a new formulation of exotic gauge field theories\cite{sub,genem,witten} and for their practical applications.  Most notably, fractons present a novel route to the possible construction of self-correcting quantum memories, providing a platform for quantum information storage and processing.\cite{haah,bravyi,terhal}  Fracton physics has also shed new light on many seemingly disconnected areas of theoretical physics, including  explaining the restricted mobility of topological crystalline defects\cite{elasticity,gromov,pai,super,potter} and providing a new mechanism for many-body localization in the absence of disorder.\cite{chamon,glassy,spread,spread2,pollmann,KhemaniScars} Fractons have also drawn a bevy of unexpected connections with topological order\cite{fracton1,fracton2,generic,fusion}, quantum Hall physics\cite{witten,phases}, deconfined quantum criticality\cite{deconfined}, gravitation\cite{mach} and holography\cite{holo}.  We refer the reader to Reference \onlinecite{review} for a review of fractons, and to selected literature\cite{field,hanlayer,sagarlayer,entanglement,
fractalsym,cheng,algebra,lego,albert,pinch,
structure,cage,twisted1,twisted2,smn,rank2ice,younonabel,compactify,taige} for details.

\begin{figure*}[!htb]
    \centering
    \includegraphics[scale=0.16]{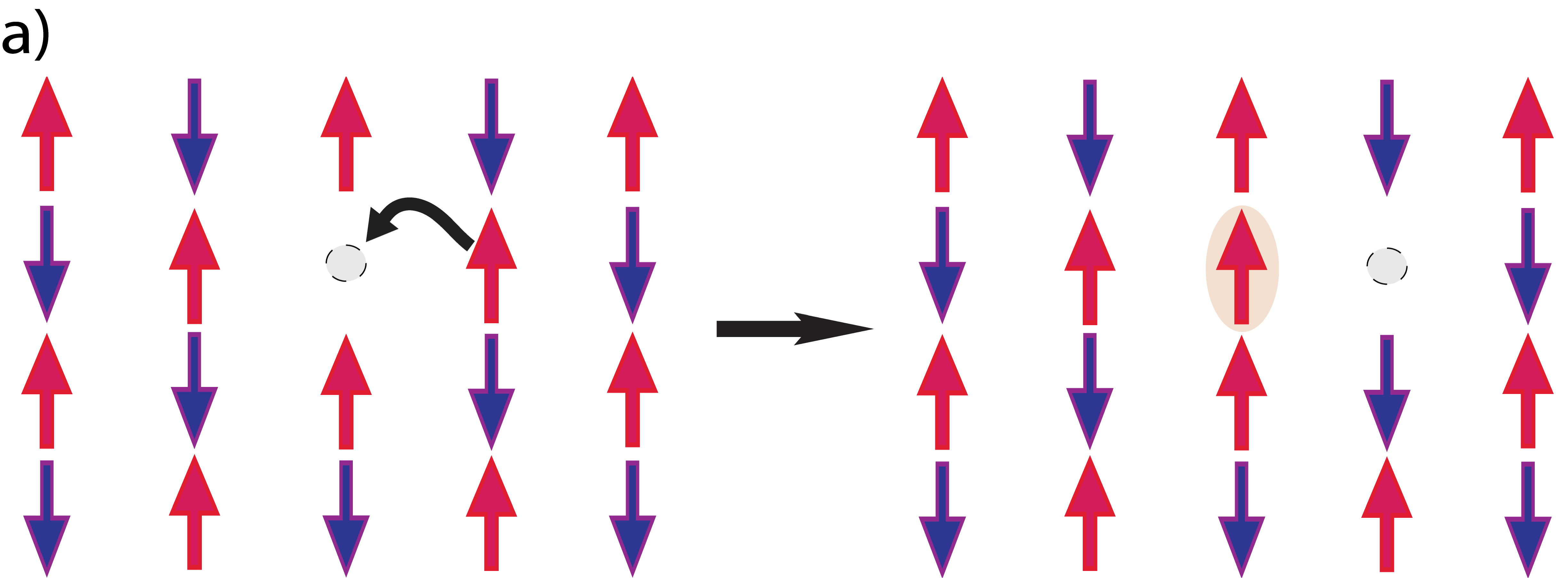} \hspace{3mm}  \includegraphics[scale=0.16]{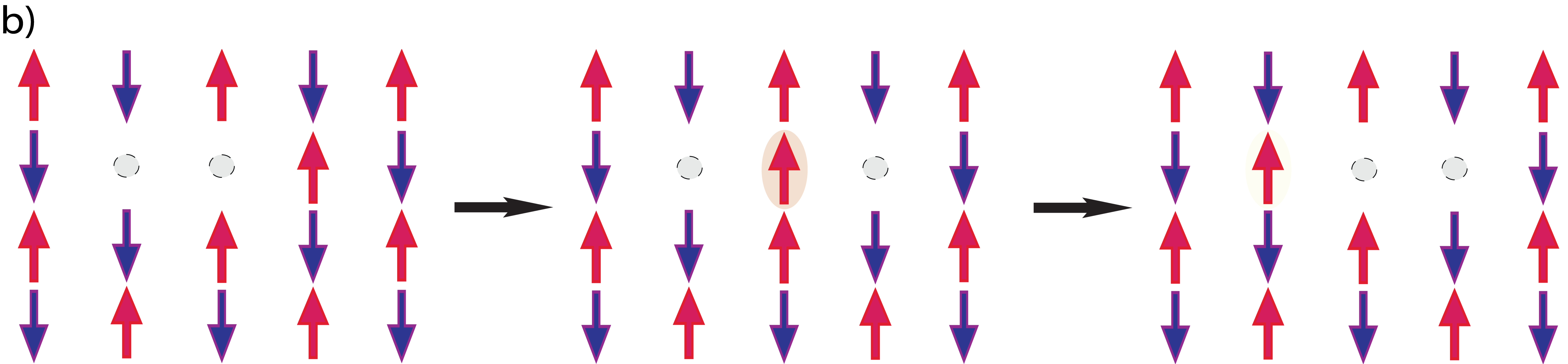}
 \caption{{\bf{Fractons from hole-doped Ising antiferromagnets.}} Motion of a hole in an Ising antiferromagnetic background is impeded due to the creation of energetically costly spin misalignments, {\em i.e.} magnons, panel a). While not perfectly immobile (due to high-order Trugman loops), a single hole is drastically less mobile than a bound state of two holes, which moves comparatively much easier, panel b). In the mixed-dimensional limit of the holes moving along a line in the two-dimensional system, the magnetic polaron ({\em i.e.} the hole dressed by magnons) is perfectly localized, as the bosonic strings restrict the hole to its original site, and Trugman loops are absent.}
 \label{fig:neel}
 \end{figure*}

Fractons were first proposed in the context of three-dimensional quantum spin-liquid models\cite{chamon,haah,fracton1,fracton2}, and were later shown to be realized as topological lattice defects of ordinary two-dimensional crystals.\cite{elasticity}  Fracton models can in principle be engineered directly using Majorana islands.\cite{lego}  But, there is still an important need for concrete, physically accessible systems realizing fracton physics.  It would be particularly useful to construct models that are realizable in one dimension, as opposed to previous models that are stable only in higher dimensions.  Such a fracton system could then be studied using the wide variety of analytic and numerical techniques available for studying one-dimensional models.

Towards this end, we here show that fracton physics can be realized in a class of models featuring boson-affected hopping (depicted in Figure~\ref{fig:polaron}), which can be found in both one- and higher-dimensional systems, see \cite{PRL} for a concise demonstration.  Systems that are described by such models include hole-doped antiferromagnets in two or three dimensions and polaronic systems in any dimension.  Such systems can exhibit either exact or approximate fracton behavior, depending on the specific details.  In these models, quasiparticles can only hop via the creation or absorption of background bosonic excitations.  For example, in a two- or three-dimensional Ising antiferromagnet, a hole can only move to a neighboring site at the expense of creating energetically costly spin misalignments ($i.e.$ magnons)\cite{Trugmanloops,ShraimanHole,KaneHole,SachdevHole,ChernHole}, as illustrated in Figure~\ref{fig:neel}.  As such, there is no ``bare'' nearest-neighbor hopping term for the holes.  Rather, a hole can only move through very weak beyond-nearest-neighbor hopping processes of a perfectly oriented spin (up arrow in Figure~\ref{fig:neel}) or through a complicated sequence of nearest-neighbor boson-mediated hoppings.\cite{Trugmanloops}

As such, the hole acquires a finite effective mass only at sixth and higher orders in perturbation theory in the nearest-neighbor limit.\cite{Trugmanloops}  In contrast, a bound state of two holes\cite{ShraimanTwoHole,DimashTwoHole,VidmarTwoHole} is capable of moving through a much simpler second-order process.  As a consequence, a broad parameter regime exists where the individual holes are effectively immobile\cite{Wohfield} compared to a ``dipole'' of two holes, which has an enormously smaller effective mass, thereby providing an approximate realization of fractons.  This intuition gained from the Ising antiferromagnet holds much more generally.  In a wide class of boson-affected hopping models, bound pairs of particles, $i.e.$ bipolarons, have significantly enhanced mobility compared to the individual particles, in close analogy with the physics of fractons.  While this fracton behavior is generically approximate, we will find a special class of boson-affected hopping models that exhibit true fracton behavior, valid to all orders in perturbation theory.

In this work, we establish a more precise relationship between fracton physics and boson-affected hopping models, such as those encountered in the study of polarons and hole-doped Ising antiferromagnets \cite{PRL}.  We begin by briefly reviewing the physics of fractons, such as their higher-moment conservation laws.  We also write a one-dimensional lattice Hamiltonian governed by such a conservation law.  This Hamiltonian features only pair-wise hopping, without single-body hopping terms, and manifestly exhibits the fracton phenomenon.  In a conventional system of particles on a lattice, such a Hamiltonian with no single-particle hopping would be enormously fine-tuned and is likely hard to engineer. However, we show that the fractonic properties of this Hamiltonian can be realized naturally (albeit approximately) in boson-affected hopping systems.  To this end, we work with a generic boson-affected hopping Hamiltonian that can describe a variety of physical systems.  We show that, upon perturbatively integrating out the bosons mediating the hopping, one naturally obtains a fracton Hamiltonian through five orders of perturbation theory.   We then move on to a particular quasi-one-dimensional boson-affected hopping model, describing holes restricted to move along one dimension of a two-dimensional system, which can be  readily realized in ultracold atoms.  We show that a hard-core constraint found in this model eliminates single-particle hopping to all orders in perturbation theory, resulting in {\em perfect} fracton physics exhibiting exact conservation of dipole moment. We note that the fracton realizations we discuss here are both ungauged and of type-I models, known to host mobile bound states. 

In this sense, systems with boson-affected hopping serve as a natural playground for explicitly studying fractonic physics, allowing to study the dynamics of fractons and dipoles as well as some of their established phenomenology, including their restrictions on thermalization and their gravitational behavior.  We investigate to what extent this phenomenology survives in approximate fracton systems, where the mobility constraints are weakly violated.  Specifically, we consider how the thermalization and gravitational properties associated with fractons are altered in the presence of a small fracton mobility, finding that both survive to an extent, providing useful experimental diagnostics of fracton physics. In fact, gravitation manifests as phase separation of holes doped in antiferromagnets, in agreement with previous studies. Besides providing a new experimentally accessible platform for fractons, boson-affected hopping systems also come with a well-established literature, on topics ranging from polarons to antiferromagnetism, from which we hope important insights may be drawn for better understanding the physics of fractons.  Our work also opens the door for application of powerful one-dimensional analytic and numerical techniques to fracton systems.

\section{The Physics of Fractons}

The essential physics of fractons is governed by perfectly correlated hopping, in which a particle can only move if there is corresponding motion of a second particle (or group of particles).  As such, in the absence of other particles with which to correlate its motion, a single fracton is strictly immobile.  Mathematically, this structure is neatly encoded in the language of higher-moment conservation laws, such as conservation of dipole moment, which severely restrict the motion of charges.  These conservation laws can arise, for example, in the context of symmetric tensor gauge theories, which describe a variety of fracton systems.\cite{sub,genem,higgs1,higgs2}  To illustrate the main principle, consider a tensor version of Maxwell theory, with a rank-2 symmetric tensor electric field, $E^{ij}$, with a charge density $\rho$ defined via a generalized Gauss's law:
\begin{equation}
\partial_i\partial_j E^{ij} = \rho
\end{equation}
(where all indices refer to spatial coordinates and repeated indices are summed over).  As compared with a conventional Gauss's law, this equation is notable for the presence of an unusual extra conservation law.  As in conventional electromagnetism, the total charge in the system is encoded in an appropriate electric flux through the boundary, which indicates that the total charge only changes through flux of charge through the boundary.  For a closed system, we can therefore conclude that the total charge is constant:
\begin{equation}
\int d^dx\,\rho = \textrm{constant}.
\end{equation}
Unlike conventional Maxwell theory, however, the presence of two derivatives in Gauss's law allows us to conclude that the total dipole moment in the system, $\int d^dx\,(\rho\vec{x})$, is also encoded as an electric flux through the boundary.  For a closed system, we can then conclude that the total dipole moment is also conserved:
\begin{equation}
\int d^dx\,(\rho\vec{x}) = \textrm{constant}.
\end{equation}
This extra conservation law severely restricts the motion of the charges of the theory.  An individual charge is not capable of moving at all, since motion in any direction would change the total dipole moment of the system.  In contrast, a dipolar bound state of two equal and opposite charges is free to move in any direction, provided it maintains its dipole moment, as indicated in Figure~\ref{fig:polaron}.  These arguments also extend to other types of tensor gauge theories, which can exhibit conservation of even higher charge moments.  In this work, however, we focus on fracton systems exhibiting only conservation of charge and dipole moment.

To date, the study of fractons has focused mainly on three-dimensional spin liquid models, as well as elasticity theory in two and three spatial dimensions.  In contrast, there has been little investigation into stable realizations of fracton physics in one dimension. Putting aside concerns of stability, one can construct a one-dimensional Hamiltonian exhibiting fracton physics by simply demanding the conservation of both charge and dipole moment, a task we accomplish in this work, see Sections~\ref{II}~\&~\ref{III}.  However, the absence of one-body hopping matrix elements, necessary for dipole conservation, is not a natural feature of ordinary systems of particles. In this work, we show how Hamiltonians of this type can naturally arise in both approximate and exact ways in the context of systems with boson-affected hopping. 

To construct a system with the desired charge and dipole conservation laws, it is simplest to consider two species of hard-core particles, created (destroyed) by $f^\dagger_\sigma$ ($f_\sigma$), which we regard as carrying opposite charges, $\sigma = \pm$.  These particles can have either bosonic or fermionic statistics, with little effect on the subsequent analysis (though we will later specialize to the case of fermions).  The dipole conservation law forbids single-particle hopping.  The lowest-order dipole-conserving hopping process corresponds to motion of the smallest dipole, a bound state of a positive and a negative charge separated by one lattice site, from one pair of sites to the next, as illustrated in Figure~\ref{fig:polaron} (see also Figure ~\ref{fig:2hop}). An effective Hamiltonian consistent with conservation laws thus takes the form: $H = -\epsilon_0\sum_{i,\sigma} f^\dagger_{i,\sigma}f_{i,\sigma} - t \sum_{i} \left( f^\dagger_{i+1,\sigma}f^\dagger_{i+2,-\sigma} +  f^\dagger_{i-1,\sigma}f^\dagger_{i,-\sigma} \right) f_{i+1,-\sigma}f_{i,\sigma}$. By design, this Hamiltonian exhibits fracton phenomenology, with mobile dipoles and stationary charges.  While this Hamiltonian is explicitly fractonic, the absence of single-body hopping matrix elements is not a natural feature of typical systems.  Such a restriction on free single-body motion, however, does arise in the context of boson-affected hopping, to which we turn next.

\section{Approximate Fracton Behavior from Boson-Affected Hopping}\label{II}

\subsection{The Model}

With an understanding of fractons in hand, we now turn our attention to what \emph{a priori} would seem like a completely disconnected area of physics.  We consider models where a set of quasiparticles, $f$, can only move via interaction with an auxiliary set of bosons, $b$.  Specifically, any process that hops an $f$ particle from one site to the next necessarily creates or absorbs a boson $b$.  We refer to this type of motion as boson-affected hopping, which we will see leads to fracton behavior for the $f$ particles.  The $f$ particles can in principle have either bosonic or fermionic statistics.  In this paper, we will mostly focus on the case where $f$ is fermionic, to match with the properties of the most common physical realizations, though a system with bosonic $f$ particles would have largely similar behavior.  From here on, therefore, we refer to the $f$ particles as the fermions and the $b$ particles as the bosons.

We have already described in the introduction how boson-affected hopping can arise in a hole-doped antiferromagnet in two or higher dimensions, since motion of a single hole requires the creation of energetically costly spin misalignments, $i.e.$ magnons.  Before moving on to an analysis of such a model, it is also useful to consider how this type of motion arises in a very different physical context, namely the study of polarons.  The concept of a polaron most commonly describes the motion of an electron in a polarizable crystal, in which the ions adjust their positions in order to screen the charge of itinerant electrons, as seen in Figure~\ref{fig:polaron}.  Dragging such a screening cloud of ions around the crystal leads to a dramatic increase in the effective mass of the electron bound in the cloud of phonons, the polaron.\cite{Mott}  Since the motion of a particle (electron or hole, etc...) in a crystal requires significant rearrangement of the background, one can effectively describe the dynamics of the particle as conditioned upon the creation or annihilation of distortions in the medium.  Subsequently, one more generally finds polaronic effects in a variety of systems ranging from particles in ordered phases\cite{Mpolaron1,Mpolaron2,Epolaron,SLpolaron} to impurities in ultracold gases.\cite{Cold1,Cold2} As an example, this approach was pioneered by Edwards \cite{edwards} in his eponymous model, which was used to study a variety of quantum phenomena in boson-affected systems.\cite{Edwards1,Edwards2,Edwards3,Edwards4,Edwards5,Edwards6} The Edwards model presumes that a carrier moves by creating and/or annihilating excitations in the background that can be parameterized as bosons. One can see that this model captures various features of the Holstein\cite{Holstein} and Peierls polaron\cite{Peierls,SousThesis} physics, magnetic polarons\cite{MP,GrusdtPRX} and the Falicov-Kimball model.\cite{FK}  We note that the phonon-induced modulation of the electron hopping in solids due to out-of-phase lattice distortions is described, to linear order, by more elaborate models, such as the Peierls model\cite{P1,P2,P3} (also known as the Su-Schrieffer-Heeger model\cite{SSH1,SSH2}).  Certain features of the boson-affected hopping models we discuss here carry over to the Peierls model, such as the pair hopping of bound pairs of fermions known as bipolarons\cite{SousBerciu}. However, other features, such as the heavily suppressed mobility of single particles, do not always carry over to Peierls polarons.\cite{Peierls}

With these physical contexts in mind, we now abstract to a generic boson-affected hopping model.  We first consider a model with only a single species of fermion $f$, which will demonstrate the central idea in the simplest context.  Later, we will extend the analysis to multi-species models, which is important for certain contexts and will make connection with even simpler fracton models.  We will only explicitly analyze one-dimensional models, though much of the same physics will carry over immediately to higher-dimensional systems.

Before focusing on purely boson-affected hopping, we first write down a model that has boson-affected \emph{and} conventional hopping processes for the fermions.  In one dimension, such a Hamiltonian can be written as:
\begin{align}
H =& \, - t_f\sum_{\langle i,j\rangle} f^\dagger_if_j + g \sum_{\langle i,j\rangle} f^\dagger_if_j(b^\dagger_j + b_i)\nonumber \\
& - \mu\sum_i f^\dagger_if_i + \omega_b\sum_i b^\dagger_ib_i.
\label{mod}
\end{align}
The first term represents unassisted hopping of the fermions, while the second represents boson-affected hopping. The last two terms are the chemical potential of the fermions and the energy cost to create a boson, respectively.  We generically take the bosons to be gapped, as in the cases of optical phonons and Ising magnons.  Note that we have not given the bosons any dynamics ($i.e.$ any $b$ hopping terms) on the grounds that, in typical physical realizations, the bosons are effectively static variables compared with the fermions.  For example, in the context of polarons, this corresponds to the statement that ions move enormously slower than electrons.  It is also worth noting that we have included only terms corresponding to boson creation on the departure site and boson absorption on the arrival site of the fermions, which is a typical situation in certain systems, such as the two-dimensional antiferromagnet of Figure~\ref{fig:neel} and the Edwards model.  In the Peierls model\cite{P1,P2,P3,SSH1,SSH2} describing the linear coupling of electron hopping to the lattice, $\sim (f^\dagger_i f_j + h.c. ) ({\rm X}_i - {\rm X}_j)$, where ${\rm X}_i \propto b_i^\dagger + b_i$, the particle can move by creating or annihilating bosons at {\em both} the arrival and departure sites.

Generically, the Hamiltonian of Equation \ref{mod} has fully mobile particles, as a consequence of the bare hopping $t_f$, with the boson coupling simply serving to modulate the effective mass.  However, in certain physical situations, such as hole-doped Ising antiferromagnets, all hopping processes necessarily involve the creation or absorption of bosonic defects, such that $t_f$ is rigorously zero.  In other cases, $t_f$ is not strictly zero, but it can often be negligibly small in strongly coupled systems.  As we will soon see, single-particle mobility is generated at sixth order in perturbation theory in this model anyway.  Thus, as long as $t_f$ is small compared to these sixth-order corrections, it will have no effect on the subsequent analysis.  Whether $t_f$ is rigorously zero or simply negligibly small, we will drop this term for the moment, writing the effective Hamiltonian as:
\begin{equation}
H = g \sum_{\langle i,j\rangle} f^\dagger_if_j(b^\dagger_j + b_i) -\mu\sum_if^\dagger_if_i + \omega_b\sum_i b^\dagger_ib_i.
\label{affect}
\end{equation}
While this Hamiltonian cannot be solved exactly for arbitrary density of $f$ particles, it can be usefully analyzed via perturbation theory expanding around the solvable point $g=0$, at which all particles are trivially stationary.

\subsection{Dynamics}

\begin{figure}[h!]
 \centering
 \includegraphics[width=1.05\columnwidth]{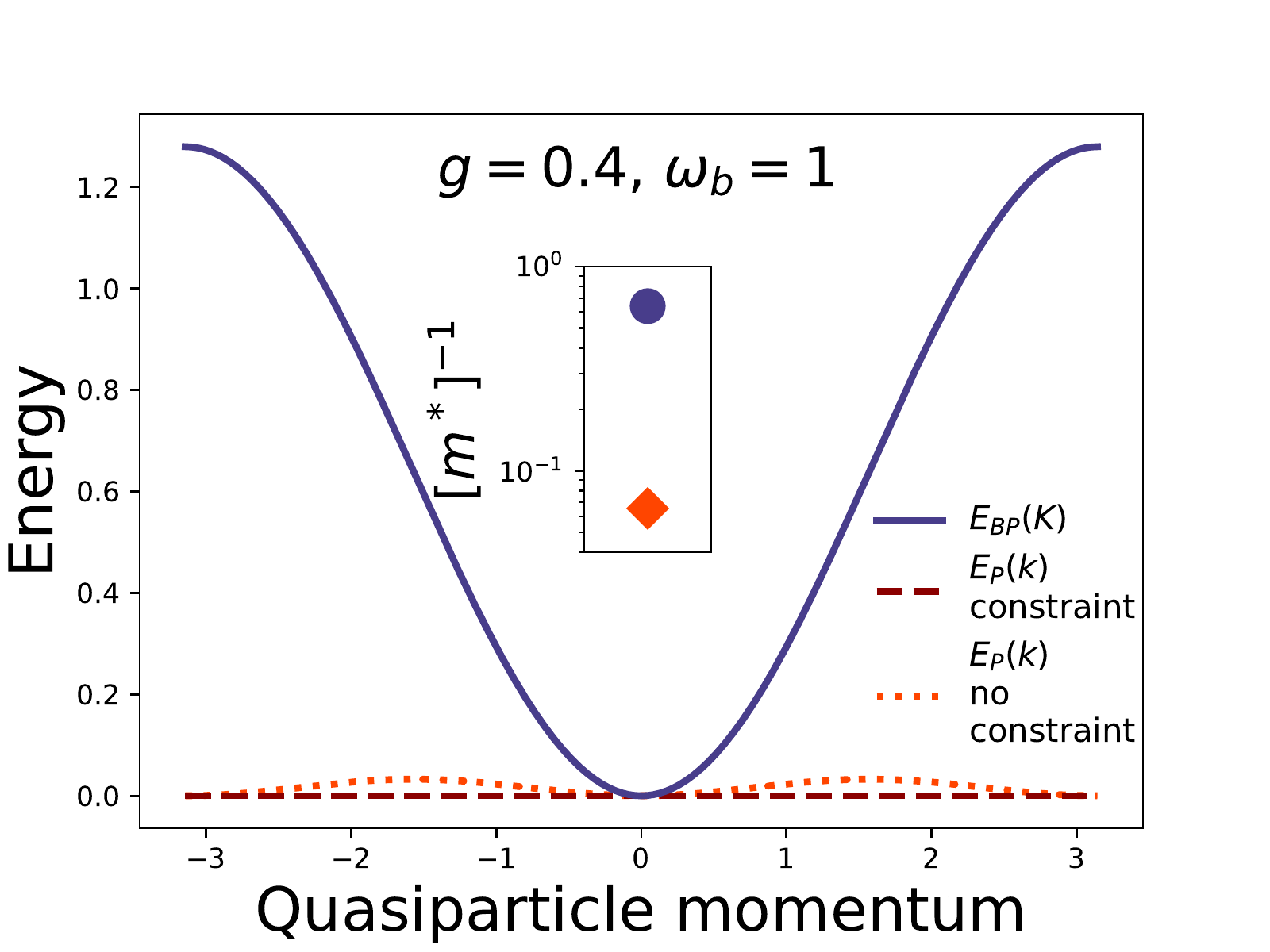}
 \caption{{\bf{Polaron (fracton) and bipolaron (dipole) dispersion and effective masses.}} We consider an exemplary case of $g=0.4$ and $\omega_b =1$ in the one-dimensional boson-affected hopping model with $t_f = 0$. To facilitate comparison, we have shifted the energy of each quasiparticle band such that the zero of energy coincides with the band minimum at the center of the Brillouin zone. Note that for two-particle states $K = k_1 + k_2$, where $k$ is the single-particle momentum. The polaron dispersion is shown for two cases: 1) in presence of a mutual hard-core constraint between the particles and the bosons (dashed line), and 2) in absence of the constraint (dotted curve).  In 1) the polaron dispersion is perfectly flat providing an exact realization of a fracton. In 2) the polaron acquires a finite bandwidth through a sixth-order process, {\em i.e.} $E_{P} (k) = -2\left(g^6/{\omega_b}^5\right)\cos(2k)$.  In the regime considered, this polaron remains much heavier than the bipolaron, whose dispersion is shown only up to second order, $E_{BP} (K)= -2\left(g^2/\omega_b\right) \cos(K)$ (solid curve).  This is further demonstrated in the inset, where we show the inverse of the masses of the polaron (diamond symbol) and bipolaron (circle symbol) in the unconstrained model. In this sense, and through staggering charge (Subsection~\ref{Staggering}), bipolarons realize dipoles, approximately in the absence of the constraint, and exactly in its presence (in one dimension).
 }
 \label{fig:motion}
 \end{figure}

\subsubsection{Single-Particle Dynamics}

\begin{figure*}[t!]
 \centering
 \includegraphics[scale=0.225]{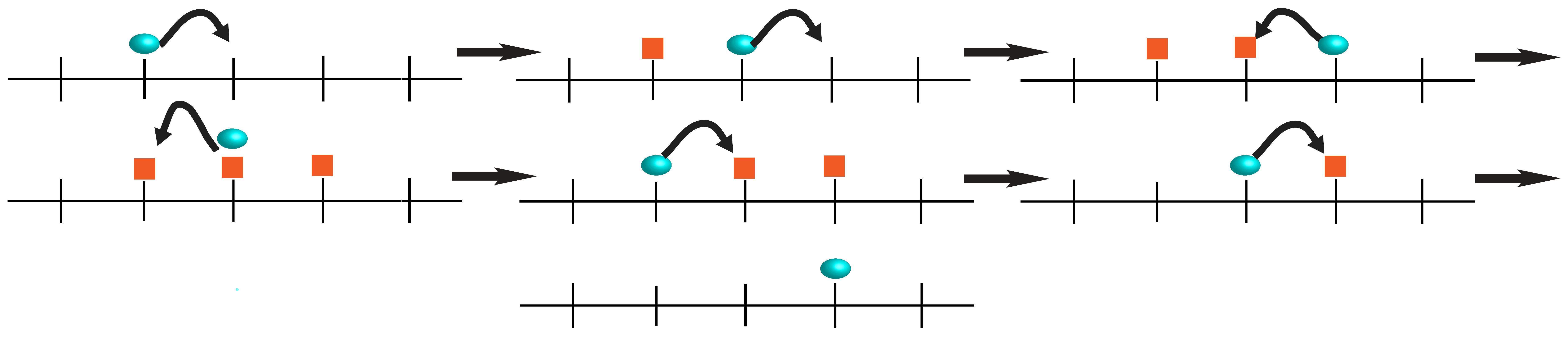}
 \caption{{\bf{Weak single-particle mobility and approximate fracton behavior.}} A schematic representation of a sixth-order process giving rise to single-particle mobility.  The particle (light-blue dot) moves by creating a string of bosonic excitations (orange-red squares), which it then ``cleans up' by retracing its steps. Through this process the particle moves two sites apart and thus acquires a $-2\left(g^6/{\omega_b}^5\right)\cos(2k)$ dispersion. As we explain in the main text, pairs of particles have a more enhanced mobility that already manifests at earlier orders, see Figure~\ref{fig:2hop}. This thus presents a case of approximately fractonic behavior.\newline} 
 \label{fig:trugman}
 \end{figure*}

We first consider the single-particle sector, satisfying $\sum_i f^\dagger_if_i = 1$ ($i.e.$ one particle in the infinite system), of the Hamiltonian in Equation \ref{affect}.  We can now find the effective Hamiltonian within this sector by performing a perturbative calculation, effectively integrating out the $b$ bosons.  The details of this perturbative calculation can be found in Appendix~\ref{AppA}.  To gain physical intuition, it is instructive to inspect the behavior at the second order. We find, to second order in $g$, that the effective single-particle Hamiltonian takes the form:
\begin{equation}
h_1 = -\epsilon_0\sum_i f^\dagger_if_i,
\label{h1}
\end{equation}
which contains only an on-site energy $\epsilon_0 = 2g^2/\omega_b$, characterizing the polaron formation energy.  (Note that this renormalization energy is half that obtained in the Peierls model of electron-phonon coupling.\cite{Peierls})  Importantly, however, the Hamiltonian does not contain any hopping processes for the fermions, so at this level of perturbation theory, the particles are strictly locked in place, behaving as fractons, see Figure~\ref{fig:motion}.  While the single-particle Hamiltonian of Equation \ref{h1} was calculated explicitly only to second order, it is easy to see pictorially that a process moving a single particle appears only at sixth order in perturbation theory\cite{Edwards4,MonaFehske2D}, see Figure~\ref{fig:trugman}.   Such processes will lead to a single-particle dispersion of order $(g^6/\omega_b^5)\cos(2k)$. Through fifth order in perturbation theory, however, the single-particle Hamiltonian will feature no hopping terms, and the particles will behave as fractons.  The polaron, which is approximately localized by the costly string excitations, thus physically realizes a fracton.

\subsubsection{Two-Particle Dynamics}

\begin{figure*}[t!]
 \centering
 \includegraphics[scale=0.235]{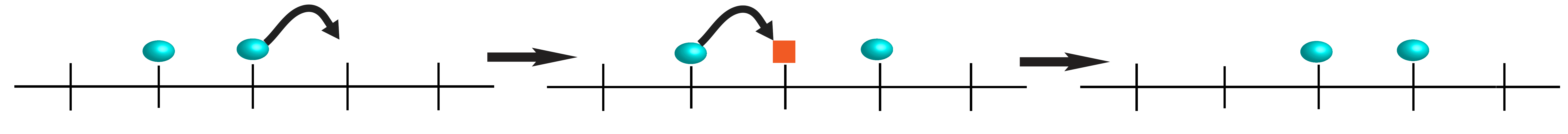}
 \caption{{\bf{Dipole-conserving two-particle dynamics.}} A two-particle bound state moves through boson-mediated pair-hopping interactions. Here, we schematically show a second-order process that involves the exchange of a single boson (orange-red square) between the two particles (light-blue dots). One particle hops first, leaving behind a boson at the site between the two particles.  The second particle then hops to the site previously occupied by the first particle by absorbing the boson.  In this way, the bound state moves over by a single lattice spacing acquiring a $-2\left(g^2/\omega_b\right)\cos(K)$ dispersion, where $K=k_1+k_2$.  Since the relative distance between the two staggered charges (see Subsection~\ref{Staggering}) remains invariant, the two-particle state provides an ideal realization of a dipole.}
 \label{fig:2hop}
 \end{figure*}

We now turn our attention to the two-particle sector of the theory, satisfying $\sum_i f^\dagger_if_i = 2$ ($i.e.$ two particles in the infinite system), which will feature nontrivial dynamics at a much lower order in perturbation theory.  Two particles become bound by moving together through a second-order process in which one particle hops and emits a boson, which is immediately absorbed by the other particle hopping in the same direction, as seen in Figure~\ref{fig:2hop}.  We again refer the reader to Appendix~\ref{AppA} for technical details of the perturbative calculation.  To second order in $g$, we obtain the effective Hamiltonian for the two-particle sector as:
\begin{align}
h_2 = -\epsilon_0\sum_i f^\dagger_if_i \,-& \,t\sum_i (f^\dagger_{i+1}f^\dagger_{i+2} + f^\dagger_{i-1}f^\dagger_i)f_{i+1}f_i \nonumber\\
&+ J\sum_i f^\dagger_if_if^\dagger_{i+1}f_{i+1},
\label{h2}
\end{align}
where $\epsilon_0 = 2t = J = 2g^2/\omega_b$.  The first term is the same on-site energy seen in the single-particle Hamiltonian, while the final term is an interaction energy between nearest-neighbor pairs.  Meanwhile, the second term represents a pair-hopping term, moving a pair of particles at sites $i$ and $i+1$ to either sites $i+1$ and $i+2$ or sites $i-1$ and $i$.  This behavior of the two-particle state mimics that of a dipole.  Importantly, however, there are still no single-particle hopping terms.  To this order in perturbation theory, we can therefore conclude that single particles are immobile, while bound states of two particles can move freely, with a dispersion of order $-t\cos(K)$ (where $K=k_1+k_2$ is the momentum of the two-polaron bound state) demonstrated in Figure~\ref{fig:motion} (see details of calculation in Appendix~\ref{AppB}), which is perfectly in line with the expected behavior of fracton systems.

Before moving on, we note that the final $J$ term of the Hamiltonian is a repulsive interaction, arising from the fact that a particle cannot hop to a neighboring site that is already occupied.\cite{SousRep,SousRepBound}  Nearest-neighbor particle pairs thereby miss out on a portion of the binding energy that would have been obtained from virtual hopping processes to neighboring sites.  Importantly, however, it should be noted that this ``repulsion'' will not destabilize the two-particle bound state.  Since the individual particles have no mobility in the Hamiltonian, they will not be able to lower their energy by moving apart.  In other words, while well-separated particles would have a lower energy than a nearest neighbor pair, there are no matrix elements in the Hamiltonian that can take the system between these two configurations.  As such, the $J$ term of the Hamiltonian merely serves to raise the energy of the two-particle bound state.

It is also important to point out that the behavior of polarons and bipolarons in such boson-affected hopping models ({\em e.g.} the Edwards model) departs from the usual view that polaronic and bipolaronic quasiparticles must be associated with mass enhancement. Indeed, here we find that single polarons experience pronounced mass enhancement. However, bipolarons do not and are light in comparison. For more discussions about light polaronic quasiparticles, we refer the reader to References \onlinecite{Peierls, SousBerciu, SousThesis} studying Peierls polarons and bipolarons, which are both shown to be light at strong coupling.  These results challenge the standard view of polaronic mass enhancement known for Holstein and Fr\"ohlich models.

\subsection{Identification of Conservation Laws} \label{Staggering}

The phenomenology of immobile particles forming mobile bound states matches perfectly with the properties of fracton systems, but to make the connection more precise, we should identify the conserved quantities in the effective Hamiltonian obtained via perturbation theory, to see how they relate to typical conservation laws in fracton systems, such as conservation of charge and dipole moment.  The effective Hamiltonian of Equation \ref{h2} manifestly obeys conservation of ``charge'' ($i.e.$ particle number), $\sum_i n^{(f)}_i = \textrm{constant}$.  However, making connection with conservation of dipole moment is slightly trickier.  The mobile excitations of the theory are bound states of two \emph{identical} particles, not opposite charges, and the motion of such a bound state does not conserve the naively defined dipole moment, $D = \sum_i n_i^{(f)} x_i$.  However, there is a simple workaround that allows us to obtain a dipolar conservation law.  We can simply define a new ``charge density'' as follows:
\begin{equation}
n_i' = n_i^{(f)}\exp\bigg(i\pi\sum_{j<i}n_j^{(f)}\bigg),
\end{equation}
which staggers the sign of charges from one particle to the next.  This definition automatically ensures that the mobile bound states consist of two particles of \emph{opposite} $n'$ charge.  In other words, the mobile bound states are true dipoles of the new charge density.  The pair-hopping $t$ processes of the Hamiltonian can move a dipole from one pair of sites to the next, but the magnitude of the dipole is always left unchanged.  Since the on-site and interaction energies ($i.e.$ $\epsilon_0$ and $J$ terms) of the Hamiltonian do not change the charge configuration of a state, we can therefore conclude that our effective Hamiltonian of Equation \ref{h2} exhibits conservation of dipole moment, with respect to the staggered charge density:
\begin{equation}
D' = \sum_i n_i'x_i = \textrm{constant},
\end{equation}
which can explicitly be checked to commute with the Hamiltonian, $[D',h_2] = 0$.  While this version of dipole conservation has a slightly nonlocal form in terms of the original fermions, due to the definition of $n'$ in terms of a semi-infinite string, it is every bit as effective at restricting their motion.  The two densities $n$ and $n'$ differ only by a sign, and any motion of a single fermion will change the value of $D'$, so the immobility of single particles can be understood as a direct consequence of this emergent conservation law.  This conservation law holds up to sixth order in perturbation theory, at which point it is violated by the generation of single-particle hopping.  Thus, this generic boson-affected hopping model gives rise to approximate fracton behavior, over a wide parameter regime between second and sixth orders of perturbation theory.  In a later section, we will investigate the properties of such approximate fracton systems and establish to what extent typical fracton phenomenology survives.

\section{Exact Fracton Behavior from Fermion-Boson Hard-Core Repulsion in Lower Dimensions}\label{III}

\subsection{The Model}

While the previous model featured only approximate fracton behavior, it is possible to realize fractons exactly through a small modification to the model system.  We now impose a mutual hard-core constraint between the fermions and the $b$ bosons, such that any site can host at most one total excitation.  This can be implemented at the Hamiltonian level, for example, by adding a repulsive term between bosons and fermions as:
\begin{eqnarray}\label{1DExactMod}
H &=& g \sum_{\langle i,j\rangle} f^\dagger_if_j(b^\dagger_j + b_i) -\mu\sum_if^\dagger_if_i + \omega_b\sum_i b^\dagger_ib_i\nonumber\\
&&\quad\quad+ \quad U\sum_i f^\dagger_if_ib^\dagger_ib_i
\end{eqnarray}
then taking the $U\rightarrow\infty$ limit.  In other words, we project all states of the form $f^\dagger_ib^\dagger_i|0\rangle$ out of the Hilbert space.  We discuss below how this constraint can be physically realized in a simple way in antiferromagnets.  For now, let us work out the physical consequences of this constrained Hilbert space.

\subsection{Dynamics}

The first notable consequence of the mutual hard-core constraint between bosons and fermions is that all of the ``backtracking'' higher-order processes contributing to single-particle motion, such as seen in Figure~\ref{fig:trugman}, are forbidden.  A fermion can only backtrack by reabsorbing a $b$ particle that it just emitted, perfectly retracing its steps.  The constraint-enforced, continuing motion of the particle in a certain one direction only involves first the creation and then subsequent annihilation of bosonic strings, effectively localizing the particle to its original position, while the  ``cleaning up''-type of motion seen in Figure~\ref{fig:trugman} is no longer possible.  As such, a single particle is now immobile to \emph{all} orders in perturbation theory, leading to perfect fracton behavior.  The single-particle Hamiltonian then takes the exact form:
\begin{equation}
h_1 = -\epsilon_0\sum_if^\dagger_if_i,
\end{equation}
where $\epsilon_0$ must be determined order by order in $g$.

Importantly, while the hard-core constraint makes single particles perfectly immobile, it still permits mobility of two-particle bound states, which thereby play the role of mobile dipoles of fractons.  This can easily be seen from the fact that the second-order process of Figure~\ref{fig:2hop} does not involve states with fermions and bosons on the same site at any point.  To second order in perturbation theory, the two-particle Hamiltonian takes the same form seen earlier:
\begin{align}
h_2 = -\epsilon_0\sum_i f^\dagger_if_i \,-& \,t\sum_i (f^\dagger_{i+1}f^\dagger_{i+2} + f^\dagger_{i-1}f^\dagger_i)f_{i+1}f_i \nonumber\\
&+ J\sum_i f^\dagger_if_if^\dagger_{i+1}f_{i+1}.
\end{align}
At this level, we see that the mutual hard-core constraint has no significant effect on dipole dynamics.

\begin{figure*}[t!]
 \centering
 \includegraphics[scale=0.25]{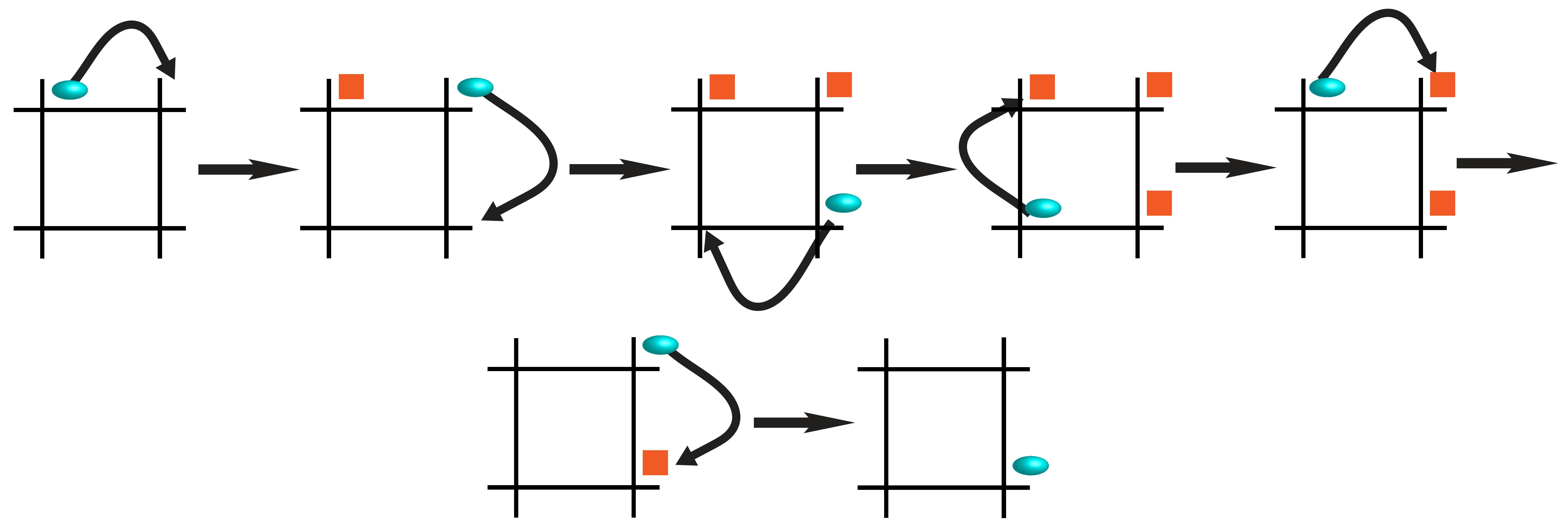}
 \caption{{\bf{Trugman loops in two-dimensional boson-affected hopping systems. }} Sixth-order processes, known as Trugman loops, involving the creation of a string of bosons (orange-red squares) around closed loops by the particle (light-blue dot), give rise to single-particle mobility, even in the presence of mutual hard-core repulsion between the particles and bosons.  If, however, the particles are restricted to move only along one direction in a two-dimensional system, as in hole-doped mixed-dimensional Ising antiferromagnets, closed loops are absent and the system is perfectly fractonic.}
 \label{fig:trugman2}
 \end{figure*}

Making use of the hard-core constraint, we can determine the dynamics of particles to even higher orders in the $g/\omega_b$ expansion.  As noted earlier, the single-particle Hamiltonian has a trivial on-site form to all orders of perturbation theory, with only the prefactor $\epsilon_0$ being renormalized order by order.  In contrast, there will be some noteworthy changes to the two-particle Hamiltonian.  To fourth order in perturbation theory, we find that the two-particle Hamiltonian takes the form:
\begin{eqnarray}\label{4thorderPT}
 h_2 &=&-\epsilon_0 \sum_{i}  n^f_{i}  +  J_{z_{1}} \sum_{i} n^f_{i} n^f_{i+1}+ J_{z_{2}}  \sum_{i} n^f_{i}n^f_{i+2} \nonumber \\  
&& \quad- \quad t_1 \sum_{i} \left( f^\dagger_{i+1} f^\dagger_{i+2} + f^\dagger_{i-1} f^\dagger_{i} \right) f_{i+1} f_{i} \nonumber\\
 && \quad + \quad t_2 \sum_{i} \left( f^\dagger_{i+2} f^\dagger_{i+3} + f^\dagger_{i-2} f^\dagger_{i-1} \right) f_{i+1} f_{i}.
\end{eqnarray}
We refer the reader to Appendix~\ref{AppA} for calculational details.  The first three terms represent an on-site energy and interactions between nearest neighbors and next nearest neighbors.  The second line is the same pair-hopping interaction we saw at second order, moving a pair of particles over by one site.  The last term is also a pair-hopping interaction that moves a pair of particles over by two lattice sites.  This new term also manifestly conserves the dipole moment $D' = \sum_i n'_ix_i$ of the staggered charge density.

More generally, let us schematically consider the form of $h_2$ to any order in perturbation theory.  It is clear that, by repeated emission of bosons by one fermion and repeated absorption by the other, a nearest-neighbor bound state of particles can hop to any location.  As such, the exact Hamiltonian, to all orders in perturbation theory, will contain terms hopping a pair by arbitrary distances.  Furthermore, the constraint forbids terms that change the net separation between the two fermions, as those require backtracking. It is also easy to check that there are no processes that can move a pair of particles that are not nearest neighbors, as there are no sufficiently long-ranged terms in the Hamiltonian.  As such, we conclude that the exact two-particle Hamiltonian takes the schematic form:
\begin{eqnarray}\label{JHam}
 h_2 &=&- \epsilon_0 \sum_{i} n^f_{i}  +   \sum_{i,\delta} {J_{z_{\delta}}} n^f_{i} n^f_{i+\delta}  \nonumber \\
&& \quad \quad  - \quad \sum_{i,\delta} t_\delta \left( f^\dagger_{i+\delta} f^\dagger_{i+\delta+1} + f^\dagger_{i-\delta} f^\dagger_{i-\delta+1} \right) f_{i+1} f_{i}, \nonumber \\
\end{eqnarray}
where $\epsilon_0$ is determined by the order of the expansion, ${J_{z_{\delta}}}$ is a density-density interaction between particles $\delta$ sites apart, and $t_\delta$ is the matrix element for hopping a pair by $\delta$ sites. Both decay rapidly as a function of $\delta$.  All terms of this form manifestly conserve the dipole moment $D'$, which we can then rigorously conclude is conserved to all orders, $i.e.$ $[D',h_2]$ is exactly zero.

\subsection{Experimental Realization: Hole-Doped Mixed-Dimensional Ising Antiferromagnets}

We have shown that a one-dimensional boson-affected hopping model supplemented by a mutual hard-core constraint between fermions and bosons exhibits perfect fracton behavior, with strictly immobile individual particles and mobile two-particle bound states.  For this to be meaningful, however, it is important to establish an experimentally realizable physical context described by such a model.  To this end, we first note that such a mutual hard-core constraint is naturally found  in the context of hole-doped antiferromagnets, since there is no meaningful way that a hole can exist on the same site as a misaligned spin.  In one dimension, a hole doped in an antiferromagnet exhibits spin-charge separation and magnons formed as misaligned spin ``defects'' can only exist in higher dimensions. In other words, holes in antiferromagnets only exhibit boson-affected hopping in dimensions higher than one.  Our proof of exact fracton behavior was, however, carried out only in one dimension.  Indeed, in two (and higher) dimensions, there are sixth-order processes known as Trugman loops leading to single-particle mobility (see Figure~\ref{fig:trugman2}), so fracton behavior is once again approximate in these higher-dimensional systems, even when supplemented by a hard-core constraint.  

Fortunately, however, there is an intermediate situation between one- and two-dimensional systems that ideally serves to realize perfect fracton behavior. In so-called ``mixed-dimensional'' Ising antiferromagnets, the hole motion is restricted to a line in a two-dimensional antiferromagnet.\cite{demler} On one hand, spin-charge separation cannot occur as the hole moves through the lattice unlike in the one-dimensional case. On the other, the holes cannot move in loops and thus cannot acquire a finite effective mass\cite{Trugmanloops}, so they behave as fractons to all orders of perturbation theory (the flat band in Figure~\ref{fig:motion}).  We refer the reader to Reference \onlinecite{PRL} for a detailed construction of fractons in hole-doped antiferromagnets.  Here, we summarize the main idea.

The mixed-dimensional Ising antiferromagnet can be potentially realized in promising experiments with Rydberg-atom arrays\cite{Labuhn,Zeiher1,Zeiher2}, trapped ions\cite{Porras, Britton}, and polar molecules\cite{Barnett,Gorshkov}, and perhaps with ultracold atoms in optical lattices.\cite{Boll, Cheuk,Mazurenko} By experimentally tuning to the limit of nearest-neighbor Ising interactions, the parent undoped system of such a two-dimensional square lattice is described by an effective Hamiltonian: $H_{\rm Ising} = J \sum_{\langle i,j\rangle } S^z_i S^z_j$. The ground state of this Hamiltonian is a classical N\'eel state $\ket{\Psi_{\rm GS}} = \Pi_{i\in A} c^\dagger_{i,\uparrow} \Pi_{j\in B} c^\dagger_{j,\downarrow}\ket{0}$ with all spins on sublattice $A$ up, and all spins on sublattice $B$ down. Here the $c^\dagger_{\uparrow/\downarrow}$ creates a particle (fermion) with spin $\uparrow/\downarrow$. Individual holes doped into this system can move through the hopping of the particles carrying the spin, which can be taken to a very good approximation, to be nearest neighbor. As outlined in Reference \onlinecite{demler}, by applying a strong gradient potential $V(y)$ along the $y$-direction taken to be one of the principal axes of the square lattice, the hole is forced to move only along the $x$-direction. We expect that mixed-dimensional antiferromagnets can also be engineered in solid-state devices.

The motion of the hole occurs as a result of the hopping of a particle whose spin becomes either perfectly oriented or disoriented with respect to the sublattice it belongs to.  We can thus regard the disoriented spin as a bosonic defect or a magnon with a creation operator:
\begin{eqnarray}
d_i^\dagger = \begin{cases}
     \sigma_i^-, & \text{if $i \in A$}.\\
     \sigma_i^+, & \text{if $i \in B$},
  \end{cases}
\end{eqnarray}
where $\sigma^\pm$ is the spin-$1/2$ raising/lowering Pauli matrix. We define the hole operator as
\begin{eqnarray}
h^\dagger_{i,\downarrow} =  c_{i,\uparrow}, & \text{if $i \in A$}.\\
h^\dagger_{i,\uparrow} =   c_{i,\downarrow}, & \text{if $i \in B$}.
\end{eqnarray}

The only way a hole can move is through the creation of a defect, which represents the displaced particle now with a misaligned spin orientation. Implementing these processes we obtain a Hamiltonian\cite{MonaFehske2D}:

\begin{eqnarray}
{
H = -t \sum_{\langle i,j \rangle,\sigma}  \left[ h^\dagger_{j,\sigma} h_{i,\sigma} (d_i^\dagger + d_j) + h.c.\right] + H_{{\rm Ising}}.
}
\end{eqnarray}
Here $\langle . \rangle$ refers to nearest neighbors, and the Hamiltonian respects a no-double occupancy constraint such that each site has either a hole or a spin: $\sum_\sigma h^\dagger_{i,\sigma} h_{i,\sigma} + d_i^\dagger d_i + d_i d_i^\dagger=1$.  

Notice that this model resembles that of Equation \ref{1DExactMod}. Here the holes take the role of the $f$ fermions, and the magnons that of the $b$ bosons. We observe that this Hamiltonian naturally respects the mutual fermion-boson hard-core constraint as a hole can only exist at a site empty of a spin.  Furthermore, in both models each site can host at most one boson. This is easy to see for the Ising antiferromagnet, for which a magnon corresponds to a spin flip of a spin-1/2 particle. For the model Equation \ref{1DExactMod}, while such a condition is not formally enforced, we note that due to the absence of backtracking, it is impossible to create more than one boson per site. Finally, we note that the bosonic excitations in the hole-doped Ising antiferromagnet cannot be described by a term such as $\omega_d \sum_i d_i^\dagger d_i$ as bosonic defects nearby cost less energy to create than those farther away. This, however, poses no qualitative difference on the physics in the limits we consider here\cite{MonaFehske2D}. We can easily see perfect fracton and dipole behavior in this system up to all orders in perturbation theory, Equation \ref{JHam}. This physics of the quasi-one-dimensional model extends to two dimensions in the low-energy limit corresponding to dynamical timescales shorter than the inverse of the sixth-order Trugman loop, promoting ideal fracton behavior to a broad parameter regime in the more widely accessible two-dimensional antiferromagnets.

In contrast to the $f$-$b$ model considered above, the holes come in two spin flavors.  Therefore, one can regard the hole's spin as a fracton's `charge' degree of freedom, then conservation of dipole moment is automatic with no need for staggering charge.\cite{PRL}

\rtext{
Besides the mixed-dimensional limit antiferromagnet, new avenues of research on approximately fractonic quasiparticles exist in two-dimensional Ising antiferromagnets. Next-nearest neighbor (spatially diagonal) hopping presents a complication since it promotes single-particle mobility. Coupling one-dimensional spin chains or Moir\'e engineering to realize rectangular antiferromagnets with diagonal distances sufficiently larger than those of square lattices presents one possible solution. Dysprosium phosphases and dysprosium aluminum garnets serve as good avenues to approximately realize our model in two dimension and in which a $t'$ term is very likely absent \cite{Wolf}.  Using cold-atom quantum simulators of mixtures of bosonic and fermionic states activated optically to realize the fermion-boson model or Rydberg simulators of hole-doped Ising antiferromagnets provide another alternative.    
}

The realization of fracton physics in polaronic systems is the central result of our work. Indeed, experiments studying magnetic polarons already show indications of the fracton phenomenon, including their restricted mobility and the string-mediated binding of dipoles.\cite{Exp1, Exp2, Exp3, Exp4, Exp5} In the next section, we suggest sharp diagnostics that will help efforts targeting the observation of the exotic features of fractons.

\section{Diagnostics and Phenomenology}

In the previous sections, we have seen how systems with boson-affected hopping can give rise to fracton physics.  We now describe ways to diagnose the presence of fracton physics in such systems, such as correlation function signatures and typical fracton phenomenology, such as restricted thermalization and gravitational behavior.  In many cases, the fracton behavior of these models is only approximate, breaking down at some high order in perturbation theory.  We therefore also study to what extent typical fracton phenomenology survives in approximate fracton systems.

\subsection{Pair Correlation Function}

One immediate consequence of the fracton behavior of boson-affected hopping models is the fact that, within the two-particle sector, the distance between those particles after integrating out the bosons never changes.  This is true regardless of whether we consider a nearest-neighbor pair, which is free to move around the system, or a stationary configuration with greater separation between the particles.  This perfect locking of the two particles with each other will have a clear manifestation in the density-density correlation functions of the system.  For example, let us consider the real-space boson-integrated density-density correlation function:
\begin{equation}
C(d) = {\rm Tr}_b  \langle \frac{1}{N} \sum_i \hat{n}_i\hat{n}_{i+d}\rangle
\end{equation}
(which is independent of $i$ for a translationally invariant system).  Here ${\rm Tr}_b$ is a trace over the $b$ particles.  Since, for any given two-particle state, the particles are separated by a constant distance $D$ (the dipole moment), this correlation function will be nonzero only for $d=D$, such that:
\begin{equation}
C(d) \sim \delta(d-D).
\end{equation}
In contrast, a two-particle state in a system without dipole conservation would feature a more generic distribution of this correlation function, without such a sharp peak.  This correlation function may prove a useful tool for detecting fracton behavior in both experimental and numerical contexts.

For contexts in which fracton behavior is approximate, not exact, the density-density correlation function will no longer be a strict delta function at a fixed dipole moment.  For dipoles with $D>1$ ($i.e.$ with particles separated by more than one lattice site) in the boson-affected models we consider, there will no longer be any reason for the two-particle configuration to have any particular fixed dipole moment in the presence of single-particle mobility, due to absence of long-range pair-hopping interactions, see Equation \ref{JHam}.  As such, the correlation function in the approximate case would eventually flatten out completely.  For $D=1$ dipoles, however, it is still energetically favorable for two particles to form a dipole, due to the gain in kinetic energy in the bound state.  While a $D=1$ state does not have its dipole preserved precisely, we still expect most of the wavefunction to have its weight in the $D=1$ sector as long as the single-particle hopping is sufficiently small.  We then expect that the system will have a rounded, but still prominent, peak in $C(d)$ near $d=1$.  This behavior can be interpreted in terms of an effective attraction between the particles of the system, a topic to which we turn next.

\subsection{Gravitational Behavior}

A hallmark feature of fracton systems is the presence of a universal attractive force between the fractons, which has a direct interpretation as an emergent gravitational force.\cite{mach,holo}  This attraction arises as a simple consequence of the fact that fractons are more mobile ($i.e.$ have a smaller effective mass) in the vicinity of other fractons.  As a toy model to illustrate the principle, consider a particle with an effective mass $m(r)$, where $r$ is the distance away from a second particle in the system, which we regard as fixed.  Neglecting any inter-particle interactions, the energy of one particle can be written as:
\begin{equation}
E = \frac{1}{2}m(r)v^2.
\end{equation}
In terms of the constant energy $E$, the velocity of the particle is then given by:
\begin{equation}
v = \sqrt{\frac{2E}{m(r)}}.
\end{equation}
Since the effective mass of a particle decreases at small $r$, the result is an increase in velocity when particles are close, and a corresponding decrease in velocity as particles move away, amounting to an effective attraction.  Note that this attraction holds whether the fracton behavior is exact ($m(r)\rightarrow\infty$ as $r\rightarrow\infty$) or is simply approximate.

We can also generalize this toy analysis to include the effects of an intrinsic short-range repulsion, as we found earlier in the Hamiltonian of Equation \ref{h2}.  In this case, the velocity of a particle can be written as:
\begin{equation}
v =\sqrt{\frac{2(E-V(r))}{m(r)}},
\end{equation}
where $V(r)$ is some short-range repulsive potential, such as $V(r) = V_0e^{-r/a}$ for lattice scale $a$.  Let us take the decrease in $m(r)$ with $r$ to be short range, which is the generic case.\cite{mach}  We model the approximate fracton behavior by allowing for a small single-particle hopping parameter $t_0$, such that the total effective hopping takes the form $t = t_0 + \eta e^{-r/a}$ for some parameter $\eta$, which sets the energy scale of the dynamics of the fracton.  Taking advantage of the fact that $t$ sets the scale of the inverse effective mass, we can write the velocity profile of the particle as:
\begin{align}
v \sim \sqrt{2t_0(E-V_0e^{-r/a})(1+\frac{\eta}{t_0}e^{-r/a})}\nonumber\\
\approx\sqrt{2t_0E}\Bigg(1+\frac{1}{2}\left(\frac{\eta}{t_0}-\frac{V_0}{E}\right)e^{-r/a}\Bigg),
\end{align}
where the last line is an approximation for $r$ bigger than a few lattice spacings (at long distances).  Importantly, note that $t_0\ll \eta$ in our situation of interest.  Specifically, in the boson-affected model we have been considering, we have $\eta \sim g^2/\omega_b$, while $t_0\sim g^6/\omega_b^5$.  As such, we can conclude that $\eta/t_0\sim (g/\omega_b)^{-4} \gg V_0/E$ for small $g$ at fixed $\omega_b$, since $V_0/E$ scales as $g^2/\omega_b$.  We can therefore identify the force between particles as attractive for almost all states.  (Note that this condition will fail for states with sufficiently small $E$, but small-$E$ states are those whose fractons are far apart such that they do not pass close to each other near $r=0$, and so they do not interact anyway).  In this way, we see that even systems with only approximate fracton behavior will exhibit a near-universal attraction between particles, which we therefore expect will cluster together in the system.  This clustering of holes will result in their phase separation\cite{PRL}, consistent with previous studies on hole-doped antiferromagnets\cite{emery,Marder,batista}, providing a smoking-gun signature of fracton behavior. (We remark that in certain physical contexts, such as electron systems, the short-range gravitational attraction between particles will compete with a long-range Coulomb repulsion (which may though be screened).  This will still lead to clustering effects at short distances, which may then be overtaken by emulsion physics at longer scales.\cite{phases})

\subsection{Restricted Thermalization}

Another notable feature of fracton systems is the fact that they tend to reach thermal equilibrium very slowly, if at all.  It has been shown that three-dimensional fracton systems approach equilibrium logarithmically slowly, in a manifestation of ``asymptotic many-body localization.''\cite{glassy}  Even more dramatically, certain one-dimensional fracton systems can fail to reach equilibrium at all, forever maintaining a memory of their initial conditions.\citep{spread}  Specifically, a system initialized with a fracton at a specific location will forever remember the position of that fracton.  A proposed explanation is based on the properties of random walks in one dimension, in which particles always eventually return to their origin.  As such, when a fracton moves by emitting a dipole, that dipole should eventually return to the fracton and be reabsorbed.  We expect similar localization of fermions in boson-affected hopping models exhibiting perfect fracton behavior, such as the mixed-dimensional Ising antiferromagnet, at least under similar initial conditions.  When a fermion moves, it is accompanied by the creation of a string of bosons.  Even if the boson has weak dynamics, it will likely return to the vicinity of the fermion and be reabsorbed, preserving localization of the fermion, at least approximately.  We leave a more rigorous analysis of these ideas to future work.  This localization, which occurs even in the absence of disorder, could provide a key signature in diagnosing fracton physics in these systems.

For approximate fracton behavior, single-particle mobility will eventually cause systems to relax to thermal equilibrium.  However, the relaxation rate will be highly dependent on the initial conditions, since two-particle bound states can disperse their energy around the system much more effectively than single particles can.  To see this, let us consider a system of particles moving at a characteristic velocity $v$.  When the system has only $\mathcal{O}(1)$ particles, the thermalization time will be limited by $\tau \sim L/v$, $i.e.$ the time it takes a particle (fracton or dipole) to travel ballistically across the system length $L$.  At larger densities, the thermalization time will be limited by diffusive processes, $\tau\sim (L^2/Dv)$, where $D$ is the diffusion constant of the system, set by the density.  In either case, the relaxation time depends inversely on the characteristic velocity $v$, which behaves roughly as $v\sim\sqrt{kT/m}$ for particles of effective mass $m$.  Since two-particle bound states have a mass $m_2\ll m_1$, where $m_1$ is the single particle mass, we see that such dipolar states will have a much lower thermalization time.  Such a parametric difference in relaxation times between states initialized with isolated particles versus two-particle bound states is a clear indication of the presence of approximate fracton physics.

Finally, we note that, due to the large separation in scale of the effective masses of fractons and dipoles, there is the possibility that dipoles may also end up localized (or at least slow to thermalize), by a mechanism akin to the quantum disentangled liquid\cite{qdl}, as we discuss further in the next section.

\section{Extensions: Multi-Species Systems and Finite Fracton Densities}

We have now established a robust connection between the physics of fractons and systems with boson-affected hopping, such as polarons.  In this section, we discuss various ways in which our analysis can be usefully extended.  We also outline several directions for future topics of investigation relevant to these ideas.

\subsection{Multi-Species Systems}

In the previous sections, we have considered boson-affected hopping models featuring only a single species of fermion, which is appropriate to certain physical contexts.  In other situations, however, the fermions can come with an internal flavor, such as the spin degree of freedom.  It is therefore important to work out how the preceding analysis extends to multi-species boson-affected hopping models.  Also, as we will see, certain types of multi-species systems will lead to realizations of more familiar fracton systems with a simpler set of conservation laws.  For concreteness, we restrict our attention to two-species models, though models with a larger number of species could also be considered without much difficulty.

We begin by considering the most straightforward multi-species generalization of the previously studied boson-affected hopping model, which we write as:
\begin{align}
H = g\sum_{\langle i,j\rangle,\sigma}f^\dagger_{i,\sigma}f_{j,\sigma}(b^\dagger_j + b_i) + \omega_b\sum_i b^\dagger_ib_i - \mu\sum_{i,\sigma}f^\dagger_{i,\sigma}f_{i,\sigma},
\label{naive}
\end{align}
where $i$ and $j$ are site indices and $\sigma$ runs over the two species, which we label as $+$ and $-$.  Once again, we can perturbatively eliminate the bosons to derive an effective Hamiltonian for the fermions.  Carrying through the same perturbative calculation as in the single-species case, through second order in perturbation theory, the effective Hamiltonian within the single-particle sector takes the form:
\begin{equation}
h_1 = -\epsilon_0 \sum_{i,\sigma} f^\dagger_{i\sigma}f_{i\sigma}
\end{equation}
with $\epsilon_0 = 2g^2/\omega_b$.  At this level, the two species behave completely independently, both described by Hamiltonians without hopping terms.  As in the single-species case, the first processes contributing to single-particle mobility occur at sixth order in perturbation theory, just as in Figure~\ref{fig:trugman}.  Below this order, both species of fermions behave as fracton excitations.

Similarly, we can calculate the effective Hamiltonian within the two-particle sector, which, to second order in perturbation theory, takes the form:
\begin{align}
h_2 =& -\epsilon_0\sum_{i,\sigma} f^\dagger_{i,\sigma}f_{i,\sigma} \nonumber\\
&- t\sum_{i,\sigma,\sigma'}(f^\dagger_{i+1,\sigma}f^\dagger_{i+2,\sigma'} + f^\dagger_{i-1,\sigma}f^\dagger_{i,\sigma'})f_{i+1,\sigma'}f_{i,\sigma}\nonumber \\
&+ J_z \sum_{i,\sigma}f^\dagger_{i,\sigma}f_{i,\sigma}f^\dagger_{i+1,\sigma}f_{i+1,\sigma} \nonumber\\
&+ J_{xy} \sum_{i,\sigma} f^\dagger_{i,-\sigma}f^\dagger_{i+1,\sigma}f_{i+1,-\sigma}f_{i,\sigma},
\end{align}
where $\epsilon_0 = 2t = J_z = J_{xy} = 2g^2/\omega_b$.  The $\epsilon_0$ term and $J_z$ term are on-site energies and nearest-neighbor interactions, respectively.  The $t$ term represents pair hopping that moves two particles (either of the same or opposite species) in the same direction by one lattice site.  The final $J_{xy}$ exchange term allows two particles of opposite species on neighboring sites to switch places by exchanging a boson. It is easy to see that, if we define the total fermion density $n_i = \sum_\sigma f^\dagger_{i,\sigma}f_{i,\sigma}$, then this quantity obeys the same conservation law as in the single-species case, $\sum_i n_i' x_i = \textrm{constant}$, where $n'_i = n_i\exp(i\pi\sum_{j<i}n_j)$.

While the above model has fracton behavior up to sixth order in perturbation theory, with an emergent dipole conservation law, it still has the same sort of mildly nonlocal form as in the single-species case, due to the string $\sum_{j<i}n_j$ used to define the sign of the charge.  In light of this, it is a useful exercise to construct a boson-affected hopping model with a purely local conservation law, regardless of whether or not the model is realistic.  To this end, we will construct a model that directly maps the two species of fermions onto positive and negative charges, with $\rho_i = n_{i+} - n_{i-}$, rather than relying on a nonlocal sign structure.  However, if we want the dipole moment of this charge density, $P = \sum_i \rho_i x_i$, to be conserved, then we need to slightly change our model from the simplest multi-species model of Equation \ref{naive}.  First of all, the $J_{xy}$ term of $h_2$, which exchanges the position of a $+$ and $-$ charge, obviously violates $P$ dipole conservation.  Thankfully, this term can be eliminated by introducing a mutual hard-core repulsion between positive and negative charges, which prevents such a switch.  Another violation of dipole conservation comes from the $\sigma=\sigma'$ piece of the $t$ term, which moves two particles of equal charge in the same direction.  This issue can also be overcome by giving a sense of directionality to the particles.  Specifically, we stipulate that a positive charge can only move right by emitting a boson and left by absorbing a boson, whereas a negative charge can only move right by absorbing a boson and left by emitting a boson.  These changes can be implemented by writing a new Hamiltonian as:
\begin{eqnarray}
H' &=& g\sum_i\big(f^\dagger_{i+1,+}f_{i,+}b^\dagger_i + f^\dagger_{i,+}f_{i+1,+}b_i \nonumber\\
&& \quad \quad  \quad \quad + \quad f^\dagger_{i+1,-}f_{i,-}b_{i+1} + f^\dagger_{i,-}f_{i+1,-}b^\dagger_{i+1}\big)\nonumber\\
&\quad + &  \omega_b\sum_i b^\dagger_ib_i - \mu\sum_{i,\sigma}f^\dagger_{i,\sigma}f_{i,\sigma} + U\sum_if^\dagger_{i,+}f^{}_{i,+}f^\dagger_{i,-}f^{}_{i,-}, \nonumber\\
\end{eqnarray}
where we take the $U\rightarrow\infty$ limit.  In this limit, the only dynamical processes allowed by this Hamiltonian involve motion of dipoles with a $(\rho, -\rho)$ charge configuration, while bound states of the same charge (if present) cannot move.  As such, this model exhibits the local conservation of dipole moment, $\sum_i\rho_ix_i=\textrm{constant}$.  This demonstrates that, as a proof of principle, completely local higher-moment conservation laws can be realized in boson-affected hopping systems.  We leave the construction of more realistic models of this form to future work.

\subsection{Finite Densities of Fractons}

In this work, we have studied boson-affected hopping models, focusing on the one- and two-particle sectors.  Such an analysis was sufficient for demonstrating the (often approximate) immobility of fractons, as well as the mobile nature of dipoles.  However, it is both important and interesting to consider what happens in sectors with a larger particle number, especially with a finite density of particles.  Some aspects of the finite-density problem can be understood as simple extensions of our work, while other questions require a more complicated analysis that we leave to future work.

As a first order of business, it is important to note that the immobility of an isolated fracton is not affected by the presence of other particles far away from the fracton under consideration.  In general, the mobility of a fracton is only affected by particles on nearby sites, and in the particular model under consideration, is only determined by the presence or absence of a nearest-neighbor particle.  A particle more than a single lattice spacing away cannot give rise to fracton mobility at any order in perturbation theory.  Similarly, a dipole cannot be prevented from moving by far-away particles.  A dipole is only blocked from motion when it comes directly into contact with another particle, due to ordinary hard-core repulsion.  Assuming that boson-mediated interactions are predominantly of the two-body type, we expect the mobility of an individual fracton or dipole to be effectively unchanged when there is a finite density of other particles in the system.

While the behavior of individual particles is roughly unchanged at finite densities, there are plenty of interesting questions to ask regarding the interactions between particles.  For example, a finite-density system will generically have a finite concentration of both dipoles, with a relatively light mass, and isolated fractons, with an infinite (or at least extremely heavy) mass.  This sort of situation is precisely what is considered in the study of ``quantum disentangled liquids''\cite{qdl}, which is a fluid made from one very heavy and one very light species of particles.  It has been argued that, while the heavy particles reach thermal equilibrium, the light particles have non-ergodic behavior due to being localized in the effective disordered landscape created by the heavy particles.  Applying this logic to the present situation, we can speculate that there are certain initial conditions for which the fractons reach thermal equilibrium, while dipoles are effectively localized on the fractons.  It remains to be seen whether or not there is a more general relationship between fracton physics and quantum disentangled liquids.

More generally, there is important work to be done in analyzing the effective Hamiltonians for one-dimensional fracton systems.  For example, we have shown that certain boson-affected hopping systems give rise to the following Hamiltonian:
\begin{eqnarray}\label{RuhEq}
H &=& -\epsilon_o\sum_i f^\dagger_if_i - t\sum_i(f^\dagger_{i+1}f^\dagger_{i+2} + f^\dagger_{i-1}f^\dagger_i)f_{i+1}f_i \nonumber\\
&&\quad +\quad J\sum_i f^\dagger_if_if^\dagger_{i+1}f_{i+1},
\end{eqnarray}
which explicitly describes the dynamics of fractons (with staggered charge) and dipoles in one spatial dimension.  One can also obtain a more conventional fracton theory with no charge staggering by working with a two-species model of the form:
\begin{eqnarray}\label{RuhEq2}
H &=& -\epsilon_0\sum_{i,\sigma} f^\dagger_{i,\sigma}f_{i,\sigma} \nonumber \\
&&- \quad t \sum_{i,\sigma} \left( f^\dagger_{i+1,\sigma}f^\dagger_{i+2,-\sigma} + f^\dagger_{i-1,\sigma}f^\dagger_{i,-\sigma} \right) f_{i+1,-\sigma}f_{i,\sigma} \nonumber\\
&&+ \quad \sum_{i,\sigma,\sigma'}V_{\sigma,\sigma'}f^\dagger_{i,\sigma}f_{i,\sigma}f^\dagger_{i+1,\sigma'}f_{i+1,\sigma'} \nonumber\\
&& +\quad U\sum_i n^f_{i,\sigma} n^f_{i,-\sigma}
\end{eqnarray}
with $U\rightarrow\infty$ limit. Here $\sigma$ and $\sigma'$ run over two possible values. What sort of phases do these Hamiltonians host, and are those phases described by a simple mean-field approach?  Does the gravitational clustering of fractons play an important role in the phase diagram of this model, such as described in Reference \onlinecite{phases}? These are important questions left to the future. 

We also note that a Hamiltonian of the form Equation \ref{RuhEq} is related to the model considered in Reference \onlinecite{ruhman}. There, it was shown that a strong pair-hopping term may lead to an gapless phase\cite{Comment} with properties very different from a phase dominated by single-particle hopping.  The boundaries between such a pair-hopping phase at its two edges and the single-particle-hopping one host a Majorana mode. This indicates that, at finite fermion densities and when sandwiched between systems dominated by single-particle hopping, this fracton model (Equation \ref{RuhEq}) should give rise to topological physics.  For example, consider a one-dimensional polaronic system characterized by strong electron-phonon coupling, such that the system is dominated by pair hopping, as discussed above.  If such a system is then coupled at its boundaries to another one with weak electron-phonon coupling, dominated by single-particle hopping, the analysis of Reference \onlinecite{ruhman} reveals that the resulting boundary must host a robust Majorana mode.  At least in this specific instance, we therefore conclude that the boundaries between a fractonic phase and a non-fractonic phase lead to topologically-protected behavior.  It is also of great interest to investigate these ideas in the two-species fracton model of Equation \ref{RuhEq2}.

It is at present unclear whether there is any deeper connection between fracton physics and topological boundary modes.  It is therefore an important task to find approximate or exact solutions to one-dimensional fracton Hamiltonians, which may yield important insights into the phases of fractonic matter and their connection with more conventional fermion theories.

\section{Conclusions}

In this work, we have shown that boson-affected hopping models, which arise in the study of electron-phonon and magnetic (holes doped into antiferromagnets) polarons, provide a physical realization of fractons, in either an exact or approximate way, depending on the details of the specific model considered.  In these models, individual quasiparticles are either perfectly or nearly immobile, since single-particle motion requires the creation of costly bosonic excitations, while bound states of these quasiparticles exhibit no such mobility restrictions, closely mirroring the physics of fractons.  More concretely, we have shown how boson-affected hopping models can be mapped explicitly onto a fracton Hamiltonian with dipole conservation via perturbatively integrating out the mediating bosons.  This effective Hamiltonian contains only pair-hopping terms, corresponding to the motion of dipoles, while all single-particle hopping elements are precisely zero.  This mapping generically holds through sixth orders in perturbation theory, so there is a wide parameter regime in which these systems exhibit approximate fracton behavior.

We have also shown how to obtain exact fracton behavior in a one-dimensional boson-affected hopping model by imposing a mutual hard-core constraint between the fermions and bosons.  Such a constraint is naturally realized in hole-doped mixed-dimensional Ising antiferromagnets, in which holes in a two-dimensional antiferromagnet are confined to a one-dimensional subspace.  In these systems, dipole moment is conserved to all orders in perturbation theory, giving rise to perfect fracton behavior.

Our work identifies boson-affected hopping systems as a new platform for studying the physics of fractons, presenting potential for observing their exotic phenomenology in experiments on polarons, such as \onlinecite{Exp1,Exp2,Exp3,Exp4,Exp5}. \rtext{When fracton physics is only approximate, we can can regard the system to correspond to a small perturbation away from a fracton phase, in an appropriate sense. Using this approach we have shown that many of the reach phenomenology of fractons survive in approximately fractonic setups.} We predict polarons will feature the universal short-range attraction characteristic of fracton systems, arising from a lowered effective mass in the presence of other fractons.  We find that this universal attraction survives in systems featuring only approximate fracton behavior, provided the violation is sufficiently weak.  This attraction leaves its mark in phase separation of holes in doped antiferromagnets. We have also predicted that boson-affected hopping systems will be slow to reach thermal equilibrium, and we have estimated the corresponding thermalization time.  We conjectured that one-dimensional models with perfect fracton behavior, such as the mixed-dimensional Ising models, will exhibit true localization, even at finite temperature.

Our results open the door for a productive exchange of ideas between a range of previously distant fields, such as fractons and polarons, and there are many interesting questions that remain to be answered.  We expect that the many powerful numerical and analytic tools available for one-dimensional systems may be productively used to study the effective fracton Hamiltonians arising in boson-affected hopping models.  How do these models behave at finite densities?  Is there some precise connection that can be made with the physics of quantum disentangled liquids?   Will the pair-hopping interactions identified in our effective fracton models lead to topological edge states? Are there results from the theory of polarons that elucidate new fracton phenomenology? These and many other questions can now be formulated in light of the connections drawn by our work.

\section*{Acknowledgments}

We acknowledge useful conversations with Mona Berciu, Jonathan Ruhman, Anton Andreev, and Pablo Sala. This material is based upon work supported by the Deutscher Akademischer Austauschdienst (DAAD) short-term grant, the Natural Sciences and Engineering Research Council of Canada (NSERC) and the Air Force Office of Scientific Research under award number FA9550-17-1-0183. J.~S. also acknowledges the hospitality of the Stewart Blusson Quantum Matter Institute at the University of British Columbia.

\appendix

\begin{appendices}

\section{Perturbative Calculations} \label{AppA}

We here detail the perturbative calculations described in the main text, which allows us to obtain effective Hamiltonians for the $f$ fermions in the one- and two-particle sectors by integrating out the $b$ bosons.  We consider the simpler case of a single species of fermions. We rewrite the Hamiltonian of Equation \ref{affect} (with $\mu=0$) as:
\begin{equation}
H = H_0 + V,
\end{equation}
where we take our unperturbed Hamiltonian as:
\begin{equation}
H_0 = \omega_b \sum_i b^\dagger_ib_i,
\end{equation}
which has all particles trivially localized.  The perturbing interaction takes the form:
\begin{equation}
V = g \sum_{\langle i,j\rangle} f^\dagger_if_j(b^\dagger_j + b_i).
\end{equation}

\begin{figure}[t!]
 \centering 
 \includegraphics[width=\columnwidth]{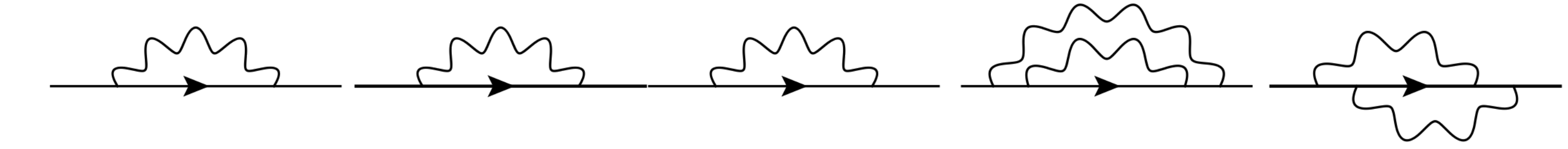} 
  \includegraphics[width=\columnwidth]{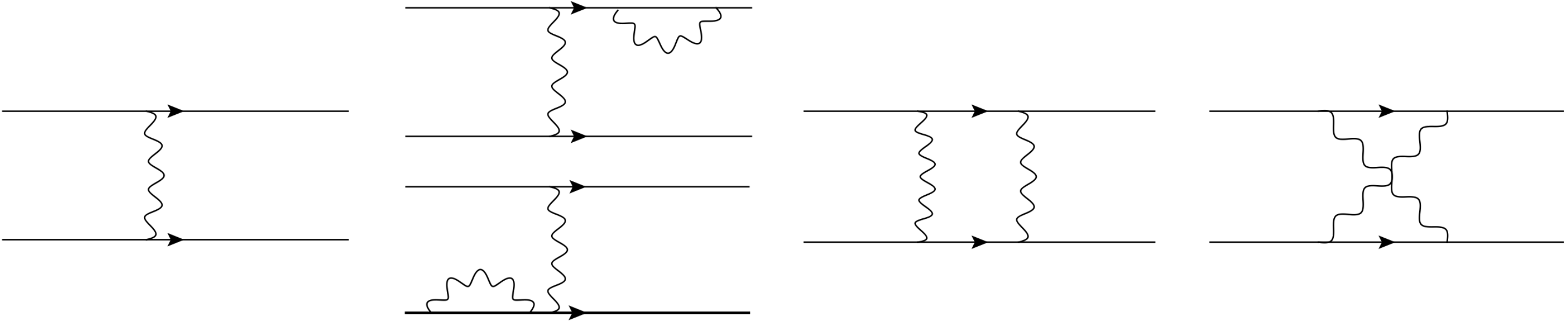}
\caption{{\bf{Feynman diagrams for one- and two-boson processes.}} A solid line with an arrow represents the particle, and a wiggly line represents the boson. Diagrams with crossed boson lines such as the last diagram in the top panel vanish in the constrained model, leading to a self-consistent solution similar in spirit to the Born approach \cite{PRL}.
}
 \label{fig::Feydiagrams}
 \end{figure}

We perform an expansion in $g/\omega_b$ integrating out the bosons at the given order of perturbation theory. It is obvious that odd orders in the expansion give zero correction. In second order, the calculation amounts to studying the effects of a single mediating boson, while in fourth order it corresponds to the effects of coupling to two bosons.  See Figure ~\ref{fig::Feydiagrams} for the set of diagrams considered in the calculation.

\subsection{Unconstrained Model}

We first  study the unconstrained model with both $f$ and $b$ particles allowed on the same site.

\subsubsection{Single-Particle Hamiltonian}
Using standard perturbation theory techniques\cite{Takahashi}, the effective single-particle Hamiltonian can be written as:
\begin{equation}
h^{{\rm 2nd}}_1 = PV\frac{1-P}{E_0-H_0}VP,
\end{equation}
where $P$ is the projector onto states with zero $b$ bosons, $E_0$ is the unperturbed energy, and $h^{{\rm 2nd}}_1$ is evaluated for single-particle states. To evaluate $VP$ on single-particle states in the Hilbert space, note that  $P$ projects out all states with non-zero bosons, and we are left with:
\begin{equation}
Vf^\dagger_i|0\rangle = g\left(f^\dagger_{i+1}+f^\dagger_{i-1}\right)b^\dagger_i|0\rangle.
\end{equation}
The $1-P$ simply returns the same state. Acting with $H^{-1}_0$ on this intermediate state simply returns the same state with an eigenvalue $\omega_b^{-1}$.  Acting with $PV$ then gives $2f^\dagger_i|0\rangle$.  Putting everything together, we have:
\begin{equation}
h^{{\rm 2nd}}_1 = -\frac{2g^2}{\omega_b}\sum_i f^\dagger_if_i,
\end{equation}
which features only an on-site energy, without any single-particle hopping terms.  As discussed in the main text, single-particle hopping only appears at sixth order in perturbation theory, as seen in Figure~\ref{fig:trugman}.

\subsubsection{Two-Particle Hamiltonian}

As in the single-particle case, the two-particle effective Hamiltonian takes the form\cite{Takahashi}:
\begin{equation}
h^{{\rm 2nd}}_2 = PV\frac{1-P}{E_0-H_0}VP
\end{equation}
evaluated on the two-particle states of the Hilbert space.  We begin by evaluating on states where the two particles are separated by one lattice site.  The projector $P$ eliminates all states with non-zero bosons, and we are left with:
\begin{eqnarray}
&&PV  \frac {1-P}{E_0 - H_0}  \left( V f_{i}^\dagger f_{i+1}^\dagger |0\rangle \right)\nonumber \\
&=& -\frac{g^2}{\omega_d} PV  \Big\{ f_{i-1}^\dagger b_i^\dagger f_{i+1}^\dagger +  f_{i} b_{i+1}^\dagger f_{i+2}^\dagger  \Big\}|0\rangle\nonumber \\
&=& -\frac{g^2}{\omega_d} \Big\{2 f_{i}^\dagger f_{i+1}^\dagger +  \left( f_{i+1}^\dagger f_{i+2}^\dagger  +  f_{i-1}^\dagger  f_{i}^\dagger \right)   \Big\} |0\rangle \nonumber \\
&=&-\frac{2g^2}{\omega_d} f_{i}^\dagger f_{i+1}^\dagger |0\rangle - \frac{g^2}{\omega_d}   \left( f_{i+1}^\dagger f_{i+2}^\dagger  +  f_{i-1}^\dagger  f_{i}^\dagger \right) |0\rangle   \nonumber \\
&\equiv&-\epsilon_0 f_{i}^\dagger f_{i+1}^\dagger |0\rangle  - t \left( f_{i+1}^\dagger f_{i+2}^\dagger  +  f_{i-1}^\dagger  f_{i}^\dagger \right)  |0\rangle.
\end{eqnarray}
The first term represents the polaron formation energy for two particles, $-2\epsilon_0$, pushed up by a nearest-neighbor repulsion $J = \epsilon_0$, while the second term is a pair-hopping interaction mediated by the bosons. 

We can also evaluate $h^{{\rm 2nd}}_2$ for the states $f^\dagger_if^\dagger_{i+n}$ for $n>1$.  In this case, the particles are sufficiently far apart that no single-boson process can allow interaction between them.  This simply generates the polaron renormalization of the energy, as expected, but no extra interactions.

Putting all these pieces together, we arrive at an effective two-particle Hamiltonian of the form:
\begin{eqnarray}
h^{{\rm 2nd}}_2 &=& -\epsilon_0\sum_i f^\dagger_i f_i -t \sum_i \left(f^\dagger_{i+1}f^\dagger_{i+2} + f^\dagger_{i-1}f^\dagger_i\right) f_{i+1}f_i \nonumber \\
&&\quad \quad + J\sum_i f^\dagger_if_if^\dagger_{i+1}f_{i+1},
\end{eqnarray}
which is Equation \ref{h2} of the main text. This Hamiltonian features only two-body hopping processes, while single-particle motion is absent (up to sixth order in perturbation theory), leading to approximate conservation of dipole moment and fracton phenomenology, as discussed earlier.

\subsection{Model with Mutual Hard-Core Constraint}

As described in the main text, we can eliminate all contributions to single-particle mobility, to all orders in perturbation theory, by imposing a mutual hard-core constraint between the fermions and the bosons of the theory that forbids the cleaning-up backtracking motion of a single fermion, such as that of Figure~\ref{fig:trugman}. While this constraint makes single particles fully immobile, it still permits the mobility of two-particle bound states.  Indeed, since the second-order perturbation theory analysis in the previous subsubsection did not involve any states violating the mutual hard-core condition, $h_2$ is identical with or without this condition, up to second order.  However, the hard-core condition allows a simplified analysis of higher-order corrections to this Hamiltonian, which we now calculate to fourth order.

\subsubsection{Single-Particle Hamiltonian}
The fourth-order correction to the effective single-particle Hamiltonian is given by\cite{Takahashi}:
\begin{eqnarray}\label{PTform}
h^{\rm 4th}_1 &=& PV\frac{1-P}{E_0-H_0}V\frac{1-P}{E_0-H_0}V\frac{1-P}{E_0-H_0}VP \nonumber \\
&& - \frac{1}{2} \Bigg( PV\left(\frac{1-P}{E_0-H_0}\right)^2VPV\frac{1-P}{E_0-H_0}VP \nonumber \\
&& \quad \quad \quad + \quad PV\frac{1-P}{E_0-H_0}VPV\left(\frac{1-P}{E_0-H_0}\right)^2VP \Bigg). \nonumber \\
\end{eqnarray}
Evaluating $h^{\rm 4th}_1$ on the single-particle states, we find
\begin{equation}
h^{{\rm 4th}}_1 = \frac{3g^4}{{\omega_b}^3}\sum_i f^\dagger_if_i,
\end{equation}
which reflects two types of polaronic renormalization processes: $f_{i}^\dagger \ket{0} \xrightarrow[]{V} b_i^\dagger f_{i\pm1}^\dagger\ket{0}  \xrightarrow[]{V} f_{i}^\dagger \ket{0} \xrightarrow[]{V} b_i^\dagger f_{i\pm1}^\dagger\ket{0}  \xrightarrow[]{V} f_{i}^\dagger \ket{0}$ and $f_{i}^\dagger \ket{0} \xrightarrow[]{V} b_i^\dagger f_{i\pm1}^\dagger\ket{0}  \xrightarrow[]{V} b_i^\dagger b_{i\pm1}^\dagger f_{i\pm2}^\dagger\ket{0}  \xrightarrow[]{V} b_i^\dagger f_{i\pm1}^\dagger\ket{0}  \xrightarrow[]{V} f_{i}^\dagger \ket{0}$. The calculation formula, Equation \ref{PTform}, keeps track of these different contributions. Note that the second type of contributions corresponds to processes in which the fermion creates and subsequently absorbs longer two-site strings of bosons.

Summarizing, we see that at the fourth order in perturbation theory the polaron formation energy is 
\begin{equation}
\epsilon_0=\frac{2g^2}{\omega_b} -\frac{3g^4}{{\omega_b}^3}.
\end{equation}

\subsubsection{Two-Particle Hamiltonian}

We now turn to the fourth-order correction to the effective two-particle Hamiltonian given by\cite{Takahashi}:
\begin{eqnarray}
h^{\rm 4th}_2 &=& PV\frac{1-P}{E_0-H_0}V\frac{1-P}{E_0-H_0}V\frac{1-P}{E_0-H_0}VP \nonumber \\
&& - \frac{1}{2} \Bigg( PV\left(\frac{1-P}{E_0-H_0}\right)^2VPV\frac{1-P}{E_0-H_0}VP \nonumber \\
&& \quad \quad \quad + \quad PV\frac{1-P}{E_0-H_0}VPV\left(\frac{1-P}{E_0-H_0}\right)^2VP \Bigg). \nonumber \\
\end{eqnarray}
A lengthy calculation of the action of $h^{\rm 4th}_2$ on all two-particle states results in 
\begin{eqnarray}
h^{{\rm 4th}}_2 &=& \frac{3g^4}{{\omega_b}^3} \sum_{i}  n^f_{i}  - \frac{3g^4}{{\omega_b}^3} \sum_{i} n^f_{i} n^f_{i+1}+ \frac{3g^4}{{\omega_b}^3}  \sum_{i} n^f_{i}n^f_{i+2} \nonumber \\  
&& \quad+\quad \frac{2g^4}{{\omega_b}^3} \sum_{i} \left( f^\dagger_{i+1} f^\dagger_{i+2} + f^\dagger_{i-1} f^\dagger_{i} \right) f_{i+1} f_{i} \nonumber\\
 && \quad + \quad \frac{1}{2}\frac{g^4}{{\omega_b}^3} \sum_{i} \left( f^\dagger_{i+2} f^\dagger_{i+3} + f^\dagger_{i-2} f^\dagger_{i-1} \right) f_{i+1} f_{i}. \nonumber \\
\end{eqnarray} 
The second term is a nearest-neighbor interaction that counteracts the polaronic renormalization with an origin similar to that discussed in the last section. Note that longer-range next-nearest-neighbor density-density and pair-hopping interactions appear first at this order.  The dipole-conserving long-range pair hopping occur as a result of a process in which, for example, one fermion leaves a two-site string of bosons for its partner to absorb, allowing the pair to move over by two sites.

Putting everything together, we obtain
\begin{eqnarray}
h_2 &=& h^{{\rm 2nd}}_2 + h^{{\rm 4th}}_2 \nonumber \\
&=&-\epsilon_0 \sum_{i}  n^f_{i}  +  J_{z_{{1}}} \sum_{i} n^f_{i} n^f_{i+1}+ J_{z_{2}}  \sum_{i} n^f_{i}n^f_{i+2} \nonumber \\  
&& \quad- \quad t_1 \sum_{i} \left( f^\dagger_{i+1} f^\dagger_{i+2} + f^\dagger_{i-1} f^\dagger_{i} \right) f_{i+1} f_{i} \nonumber\\
 && \quad + \quad t_2 \sum_{i} \left( f^\dagger_{i+2} f^\dagger_{i+3} + f^\dagger_{i-2} f^\dagger_{i-1} \right) f_{i+1} f_{i}, \nonumber \\
\end{eqnarray}
which is the effective Hamiltonian of Equation \ref{4thorderPT} and the coefficients that appeared at the second order are now renormalized by additive factors $\propto g^4/{\omega_b}^3$ at this fourth order.

Crucially, as we discuss in the main text, the mutual hard-core constraint forbids cleaning-up backtracking motion, and thus always ensures bosons are created one at a site in string configurations.  This ensures perfect polaron immobility, but does not affect the string-mediated dipole-conserving bipolaron mobility.

From this, we infer the behavior to all orders of perturbation theory in Equation \ref{JHam} of the main text. We expect the physics to hold generally even in the limit of large number of bosons, as each site will still host at most a single boson and the string configurations will just simply become longer.

\section{Equation of Motion for Dipoles} \label{AppB}

In a Hamiltonian exhibiting fracton physics, such as that of Equation \ref{h2}, the individual fermions are localized (forming polarons) and have no dispersion.  However, bound states of pairs of fermions, {\em i.e.} bipolarons, have a nontrivial dispersion.  We wish to find the equation of motion\cite{MonaFewParticle} for such bound states. We do so by solving for the Green's function, $G(\omega) = \frac{1}{\omega + i\eta - H}|_{\eta\rightarrow 0^+}$.  We first take advantage of the identity $G(\omega)[\omega + i\eta - H] = 1$, evaluating its expectation value in a set of normalized basis states.  To this end, we define momentum states as:
\begin{equation}
|K,n\rangle = \frac{1}{\sqrt{N}}\sum_i e^{iK \left(R_i+n/2\right)}f^\dagger_if^\dagger_{i+n}|0\rangle
\end{equation}
with $n>0$, and $K = k_1 + k_2$ is the dipole (bipolaron) momentum.
We derive the equations of motion by evaluating:
\begin{align}
\langle K,1|G&(\omega)[\omega + i\eta - h_2)]|K,n\rangle\nonumber \\
&= g(\omega,K,n) - \langle K,1|G(\omega)h_2|K,n\rangle = \delta_{n,1},
\label{cond}
\end{align}
where we have defined $g(\omega,K,n) = \langle K,1|G(\omega)|K,n\rangle$.  To proceed, we compute the action on the states $|K,n\rangle$ of the Hamiltonian $h_2$, working out term by term:
\begin{equation}
-\epsilon_0\sum_if^\dagger_if_i|K,n\rangle = -2\epsilon_0|K,n\rangle,
\end{equation}
\begin{eqnarray}
&&-t\sum_i\left(f^\dagger_{i+1}f^\dagger_{i+2} + f^\dagger_{i-1}f^\dagger_i\right)f_{i+1}f_i|K,n\rangle  \nonumber\\
&&\quad \quad \quad \quad \quad= -2t\cos(K)|K,1\rangle \delta_{n,1},
\end{eqnarray}
and
\begin{equation}
J\sum_i f^\dagger_if_if^\dagger_{i+1}f_{i+1}|K,n\rangle = J|K,1\rangle \delta_{n,1}.
\end{equation}
Since $|K,1\rangle$ is an eigenstate of each of the terms in $h_2$, it is also an eigenstate of $h_2$, and we can immediately identify the dipole Green's function, $g(\omega,K,1)$, from Equation \ref{cond} as:
\begin{equation}
g(\omega,K,1) = \frac{1}{\omega + i\eta + 2\epsilon_0 - J + 2t\cos(K)}.
\end{equation}
The two-polaron bound state energy corresponds to the pole of $g(\omega,K,1)$, and we find the dipole (bipolaron) dispersion:
\begin{equation}
E_{BP}(K) = -2\epsilon_0 + J - 2t\cos(K)
\end{equation}
as stated in the main text.

\end{appendices}


\begin{thebibliography}{9}
\bibitem{Comment}
The strong-pairing phase is gapless in the charge sector, but gapped to single fermion excitations.
\end{thebibliography}
\end{document}